\documentclass[12pt]{article}

\usepackage{amsmath,amssymb}
\usepackage{graphicx}
\usepackage{caption}
\usepackage{color}
\usepackage{dcolumn}
\usepackage{bm}
\usepackage[numbers,super,comma,sort&compress]{natbib}
\usepackage[T1]{fontenc}
\definecolor{background-color}{gray}{0.98}

\newcommand{\ket}[1]{|#1\rangle}
\newcommand{\bra}[1]{\langle#1|}
\newcommand{\Tr}{\text{Tr}}

\title{Geometric phases in quantum information}
\author{Erik Sj\"oqvist\thanks{Department of Quantum Chemistry, Uppsala University, Box 518, 
SE-751 20 Uppsala, SWEDEN}}

\begin{document}

\maketitle

\begin{abstract} 
The rise of quantum information science has opened up a new venue for applications of the 
geometric phase (GP), as well as triggered new insights into its physical, mathematical, and 
conceptual nature.  Here, we review this development by focusing on three main themes: the 
use of GPs to perform robust quantum computation, the development of GP concepts for mixed 
quantum states, and the discovery of a new type of topological phases for entangled quantum 
systems. We delineate the theoretical development as well as describe recent experiments 
related to GPs in the context of quantum information. 
\end{abstract}

\clearpage


  \makeatletter
  \renewcommand\@biblabel[1]{#1.}
  \makeatother

\bibliographystyle{apsrev}

\renewcommand{\baselinestretch}{1.5}
\normalsize

\clearpage

\section*{\sffamily \Large INTRODUCTION} 
A cat held upside down can fall on its feet although its angular momentum vanishes 
\cite{shapere89a}. An amoeba can swim in a completely reversibel environment 
\cite{shapere89b}. How do they manage? 

The mechanism used by these cats and amoebae is the same as that used by a quantum 
state that returns to itself, but picks up a phase factor of geometric origin. The mechanism 
is called {\it holonomy} or {\it geometric phase} (GP) and describes the twisting of a 
physical quantity, such as the orientation of the cat or the phase of the 
quantum state, due to the curved geometry of the physical space where the change takes 
place \cite{batterman03}. This space is the space of shapes in the case of cats, while it is the 
quantum state space or projective Hilbert space in the case quantum states. 

If the evolution of a quantum system is induced by slowly changing some external parameters 
around a loop, the state approximately performs a loop giving rise to an adiabatic GP. This 
kind of GP was first discovered in the field of quantum chemistry by  Longuet-Higgins and 
coworkers \cite{longuet58,herzberg63,longuet75} and Stone \cite{stone76}. It is the underlying 
mechanism behind the molecular Aharonov-Bohm effect \cite{mead80a,mead80b}, which 
takes the form of a topological phase shift when a molecule reshapes slowly around a 
conical intersection point. 

Later, Berry \cite{berry84} demonstrated that the GP is a general 
consequence of the geometrical structure of the space of parameters that drive cyclic adiabatic 
evolution of energetically non-degenerate quantum states; a scenario that can be realized 
in numerous systems, such as photons \cite{tomita86}, NMR \cite{suter87}, neutrons 
\cite{bitter87}, Jahn-Teller molecules \cite{vonbusch98}, condensed matter systems 
\cite{zhang05}, and cold atoms \cite{lin09}. Simon \cite{simon83} demonstrated that 
Berry's GP can be understood as the holonomy in a Hermitian line bundle. This result has  
established the geometrical nature of the phase. 

GPs have been generalized beyond Berry's framework. Wilczek and Zee \cite{wilczek84}, 
still within the context of adiabatic evolution, pointed out that energetically degenerate 
states may aquire matrix-valued GPs when slowly changing parameters trace out a loop 
in parameter space. The matrix nature of these phases make them potentially non-commuting; 
a fact that has motivated the term non-Abelian GP or quantum holonomy for this kind of 
phase effects. Aharonov and Anandan \cite{aharonov87} removed the restriction of 
adiabatic evolution and introduced the concept of non-adiabatic GP in arbitrary cyclic 
quantum evolution. Subsequently, Anandan \cite{anandan88} proposed non-adiabatic 
non-Abelian GPs of general quantum systems.  Based on Panchartnam's work 
\cite{pancharatnam56} on polarization of classical light fields, Samuel and Bhandari 
\cite{samuel88} defined GPs for open paths in the underlying state space. 

The rise of quantum information science has opened up a new venue for application 
of the GP, namely, to use it as a tool for robust quantum information processing 
\cite{vedral02,sjoqvist08}. This development has been triggered by the work of Zanardi 
and coworkers \cite{zanardi99,pachos99,pachos01} on {\it holonomic quantum computation}, 
i.e., the idea to use non-Abelian Wilczek-Zee GPs to implement quantum gates; the logical 
transformations that build up a circuit-based quantum computation. Such implementations 
are believed to be useful to reach the error threshold, below which quantum computation 
with faulty gates can be performed. The basic reasoning behind the conjectured robustness 
is that GP is a global feature of quantum evolution being resilient to errors, such as parameter 
noise and environment-induced decoherence, which are picked up locally along the path in 
state space. 

Conversely, tools and concepts developed in quantum information theory have been used 
to broaden the concept of GP itself. Based on early work by Uhlmann \cite{uhlmann86} on 
holonomy along paths of density operators, new concepts of GPs for statistical mixtures 
of wave functions have been developed \cite{sjoqvist00a,tong04,marzlin04,chaturvedi04}. 
In relation to these {\it mixed state geometric phases}, GPs of quantum systems that interact 
with a quantum-mechanical environment have been examined in the contexts of quantum 
jumps \cite{carollo03,fuentes05}, quantum maps \cite{peixoto03,ericsson03a,kult08}, 
stochastic unravellings \cite{bassi06,sjoqvist06,buric09,pawlus10}, and the adiabatic approximation 
\cite{thunstrom05,sarandy06,oreshkov10}. Studies of GPs of two or more quantum degrees 
of freedom have led to the discovery of {\it entanglement-induced topological phases} 
\cite{milman03,oxman11,johansson12a}. This new type of 
phases have been shown to be useful to characterize quantum entanglement.  
 
The objective of this review is to describe the recent merging of ideas in quantum information 
science and the field of GP. The basic theory of Abelian and non-Abelian GPs are outlined in 
the next section. Thereafter follows the core of this review, namely, a description and overview 
of various applications of GPs in quantum information. This part is focused on three main 
themes: quantum computation, mixed quantum states, and quantum entanglement. The 
paper ends with a concluding section, which in particular contains some pertinent issues 
to examine in the future in this research area.   
 
\section*{\sffamily \Large DIFFERENT FLAVORS OF GEOMETRIC PHASE}
The general structure of the quantum geometric phase (GP) is the removal of accumulated 
local phase changes from the global phase acquired in some evolution of a quantum system 
\cite{mukunda93}: 
\begin{eqnarray}
\textrm{GP} = \textrm{Global phase} - \sum \textrm{Local phase changes} .
\label{eq:gpgeneral}
\end{eqnarray}
The resulting GP factor can be an Abelian phase factor, denoted as $e^{i\Phi [\mathcal{C}]}$, 
or a non-Abelian unitary matrix, denoted as $U [\mathcal{C}]$, depending on the context. 
The path $\mathcal{C}$ may reside in different kinds of spaces, such as projective Hilbert space 
in the case of pure states, the space of subspaces in the case of quantum gates, and the space 
of density operators in the case of mixtures of pure states. In this section, we review the basic 
theory of these Abelian and non-Abelian GPs by illustrating them in the physical context of 
pure state evolution. 

\section*{\sffamily \Large Abelian GPs}
A well-known feature of quantum evolution driven by a time-independent Hamiltonian $H$ 
is that a stationary pure quantum state, as represented by a vector $\ket{n}$ in Hilbert space 
of the system, picks up a phase factor determined by the average energy $\bra{n} H \ket{n} = E$ 
and the elapsed time $t$, i.e., 
\begin{eqnarray}
\ket{\psi (0)} = \ket{n} \rightarrow \ket{\psi (t)} = e^{-i Et/\hbar} \ket{n} . 
\end{eqnarray} 
This simple fact is not true when the state itself evolves in time as the phase then picks 
up information of the geometry of the path in state space in addition to the system's energy. 
The origin of this extra phase contribution, being the pure state GP, is the curvature 
of state space. 

To understand the appearance of a GP shift accompanying state changes, consider 
a superposition of two stationary states $\ket{n}$ and $\ket{m}$ of the form 
\begin{eqnarray} 
\ket{\psi (0)} = a\ket{n} + b\ket{m} .  
\end{eqnarray}
Let $E_n$ and $E_m$ be the corresponding energies assumed to be non-degenerate 
and ordered as $E_m > E_n$. The coefficients $a$ and $b$ are non-zero complex numbers 
whose values determine the initial state of the system, such that $|a|^2$ and $|b|^2$ are 
the probabilities to find the energies $E_n$ and $E_m$, respectively.  Linearity of the 
Schr\"odinger equation implies that $\ket{\psi (0)}$ evolves into 
\begin{eqnarray}
\ket{\psi (t)} = a e^{-i E_n t/\hbar} \ket{n} + b e^{-i E_m t/\hbar} \ket{m} .    
\end{eqnarray} 
This describes a non-trivial cyclic evolution of the quantum state as can be seen by evaluating 
$\ket{\psi (t)}$ at time $\tau = 2\pi \hbar /(E_m - E_n)$:
\begin{eqnarray}
\ket{\psi (\tau)} & = & 
e^{-i E_n t/\hbar} \left. \left( a \ket{n} + b e^{-i (E_m - E_n) t/\hbar} \ket{m} \right) 
\right|_{t=\tau} 
\nonumber \\ 
 & = & e^{-i 2\pi E_n/(E_m-E_n)} \ket{\psi (0)} \equiv e^{if} \ket{\psi (0)} .  
\end{eqnarray}  
$\tau$ is thus the period of the cyclic evolution that corresponds to the loop 
$\mathcal{C} : [0,\tau] \ni t \rightarrow \ket{\psi (t)} \bra{\psi (t)}$ in state space. 

The resulting overall phase $f=-2\pi E_n/(E_m-E_n)$ can be divided into a sum of two parts. 
The first part is the phase induced by the average energy $\bra{\psi (t)} H \ket{\psi (t)} = 
|a|^2 E_n + |b|^2 E_m$ and reads 
\begin{eqnarray} 
\delta = - \frac{2\pi}{E_m-E_n} \left( |a|^2 E_n + |b|^2 E_m \right)  
\end{eqnarray}
in analogy with the above phase shift $Et/\hbar$ for a stationary state. $\delta$ is 
clearly energy-dependent and therefore a dynamical phase associated with the evolution. 
The remainder $\gamma = f - \delta$ reads 
\begin{eqnarray} 
\gamma = - \frac{2\pi E_n}{E_m-E_n} + \frac{2\pi}{E_m-E_n} \left( |a|^2 E_n + |b|^2 E_m \right) = 
2\pi |b|^2 , 
\label{eq:se}
\end{eqnarray}
where we have used normalization $|a|^2 + |b|^2 = 1$. Remarkably, $\gamma$ is independent 
of the energies of the system; in fact, $\gamma$ is the GP $\Phi [\mathcal{C}]$ associated with 
the loop $\mathcal{C}$. 

If we remove the second phase factor $e^{-i E_m t/\hbar}$ instead of the phase factor 
$e^{-i E_n t/\hbar}$, then the overall phase $\tilde{f} = -2\pi E_m/(E_m-E_n)$ and the 
GP $\tilde{\gamma} = -2\pi |a|^2 \neq \gamma$. By using the normalization of $\ket{\psi (0)}$, 
we find $\gamma - \tilde{\gamma} = 2\pi$, which implies $e^{i\tilde{\gamma}} = 
e^{i\gamma}$. In other words, the GP factor is the same for the two arbitary phase choices. 
Indeed, it is not difficult to prove that it is invariant under any phase choice of the evolving 
state, which makes the GP of a pure state an invariant under the Abelian group U(1), i.e., the 
group of phase transformations. 

The independence of energy and local phase choice suggest that the Abelian GP can be defined 
without reference to the underlying dynamical mechanism that drives the evolution. This can be 
seen by using the concept of relative phase \cite{pancharatnam56} between state vectors. Let 
$\ket{\psi (t_1)}$ and $\ket{\psi (t_2)}$ be two non-orthogonal vectors along a continuous 
curve $C: t \in [0,\tau] \rightarrow \ket{\psi (t)}$ in Hilbert space. The phase of $\ket{\psi (t_2)}$ 
relative $\ket{\psi (t_1)}$ is simply the phase of the scalar product $\langle \psi (t_1) 
\ket{\psi (t_2)}$. The relative phase concept can be used to define 
\begin{eqnarray} 
\textrm{Global phase} = \arg \langle \psi (0) \ket{\psi (\tau)} , 
\label{eq:globalabelian}
\end{eqnarray} 
being well-defined if $\langle \psi (0) \ket{\psi (\tau)} \neq 0$ (non-orthognal end-points), 
and 
\begin{eqnarray} 
\sum \textrm{Local phase changes} = 
\lim_{\delta t \rightarrow 0} \int_0^{\tau} \arg \langle \psi (t) \ket{\psi (t+\delta t)} = 
-i \int_0^{\tau} \langle \psi (t) \ket{\dot{\psi} (t)} dt , 
\label{eq:localabelian}
\end{eqnarray} 
where we have used Taylor expansion $\langle \psi (t) \ket{\psi (t+\delta t)} \approx 
1+\langle \psi (t) \ket{\dot{\psi} (t)} \delta t \approx e^{\langle \psi (t) \ket{\dot{\psi} (t)} 
\delta t}$ and the fact that $\langle \psi (t) \ket{\dot{\psi} (t)}$ is purely imaginary due to 
normalization: $1 = \langle \psi (t) \ket{\psi (t)} \Rightarrow 0 = \langle \dot{\psi} (t) 
\ket{\psi (t)} + \langle \psi (t) \ket{\dot{\psi} (t)} = 2 \textrm{Re} \langle \psi (t) \ket{\dot{\psi} (t)}$. 
By combining Eqs.~(\ref{eq:globalabelian}) and (\ref{eq:localabelian}) with Eq.~(\ref{eq:gpgeneral}), 
we obtain \cite{mukunda93}  
\begin{eqnarray} 
\Phi [\mathcal{C}] = \arg \langle \psi (0) \ket{\psi (\tau)} 
+ i \int_0^{\tau} \langle \psi (t) \ket{\dot{\psi} (t)} dt . 
\label{eq:abeliangp}
\end{eqnarray} 
This GP is invariant under local phase changes $\ket{\psi (t)} \rightarrow e^{i\alpha (t)} 
\ket{\psi (t)}$ and invariant under reparametrizations $t \rightarrow \tau (t)$ such that 
$\dot{\tau} > 0,  \forall t \in [0,\tau]$. Thus, GP is a property of the path $\mathcal{C}$ 
in state space, defined by the projection $\ket{\psi (t)} \rightarrow \ket{\psi (t)} \bra{\psi (t)}$ 
of the path $C$ in Hilbert space. Note that this definition of GP makes no reference to 
cyclic evolution and therefore applies to any evolution of a pure quantum state connecting 
two non-orthogonal end-points $\ket{\psi (0)}$ and $\ket{\psi (\tau)}$. 

There are three `canonical' choices of local phase: 
\begin{itemize}
\item[(i)] If $\ket{\psi (t)}$ is a solution of the Schr\"odinger equation, i.e., satisfies 
$i\hbar \ket{\dot{\psi} (t)} = H(t) \ket{\psi (t)}$, then we obtain from Eq. (\ref{eq:abeliangp}) 
the expression \cite{aharonov87}
\begin{eqnarray}
\Phi [\mathcal{C}] = \arg \langle \psi (0) \ket{\psi (\tau)} - \frac{1}{\hbar} 
\int_0^{\tau} \bra{\psi (t)} H(t) \ket{\psi (t)} dt ,  
\end{eqnarray}
where $H(t)$ is the Hamiltonian of the system. Thus, the GP can be understood as the 
removal of the accumulated phase induced by the average energy $\bra{\psi (t)} H(t) 
\ket{\psi (t)}$, i.e., the dynamical phase shift of the system, from the acquired global 
phase, just as in the example of the two superposed stationary states discussed above. 
\item[(ii)] If 
\begin{eqnarray} 
\langle \psi (t) \ket{\psi (t+\delta t)} > 0 \Rightarrow \langle \psi (t) \ket{\dot{\psi} (t)} = 0, 
\ t\in [0,\tau) 
\label{eq:pt}
\end{eqnarray}
no phase changes are acquired along the path, and we obtain 
\begin{eqnarray}
\Phi [\mathcal{C}] = \arg \langle \psi (0) \ket{\psi (\tau)} . 
\end{eqnarray}
When Eq. (\ref{eq:pt}) is satisfied, $\ket{\psi (t)}$ is said to be parallel transported. In this 
case, the GP coincides with the global phase associated with the evolution. Parallel transport 
is particularly useful in experiments as it makes the GP directly accessible by observing the 
global phase \cite{wagh95a,wagh95b,wagh98}. 
\item[(iii)] The phase choice $\ket{\lambda (t)} = e^{-i \arg \langle \psi (0) \ket{\psi (t)}} 
\ket{\psi (t)}$ is a gauge invariant reference section \cite{pati95a,pati95b} as $\ket{\lambda (t)}$ 
is unchanged under local phase transformations of $\ket{\psi (t)}$. (To be precise, 
$\ket{\lambda (t)} \rightarrow e^{i\alpha (0)} \ket{\lambda (t)}$ when $\ket{\psi} \rightarrow 
e^{i\alpha (t)} \ket{\psi (t)}$.) We note that $\arg \langle \lambda (0) \ket{\lambda (\tau)}$ 
vanishes so that 
\begin{eqnarray} 
\Phi [\mathcal{C}] = i \int_0^{\tau} \langle \lambda (t) \ket{\dot{\lambda} (t)} dt , 
\end{eqnarray}  
which is the familiar form of the geometric phase being the closed path integral of the 
effective `vector potential' $i \langle \lambda \ket{d\lambda}$. This perspective was 
adopted in Berry's original work on GP \cite{berry84}. 
\end{itemize}
Note that although the three canonical phase choices focus on different aspects of quantum 
evolution, they differ only by local phase choices and therefore all give the same numerical result for 
the GP.  

\subsection*{\sffamily \large Example: GP of a qubit}
\begin{figure*}[ht]
\centering
\includegraphics[width=1.0\textwidth]{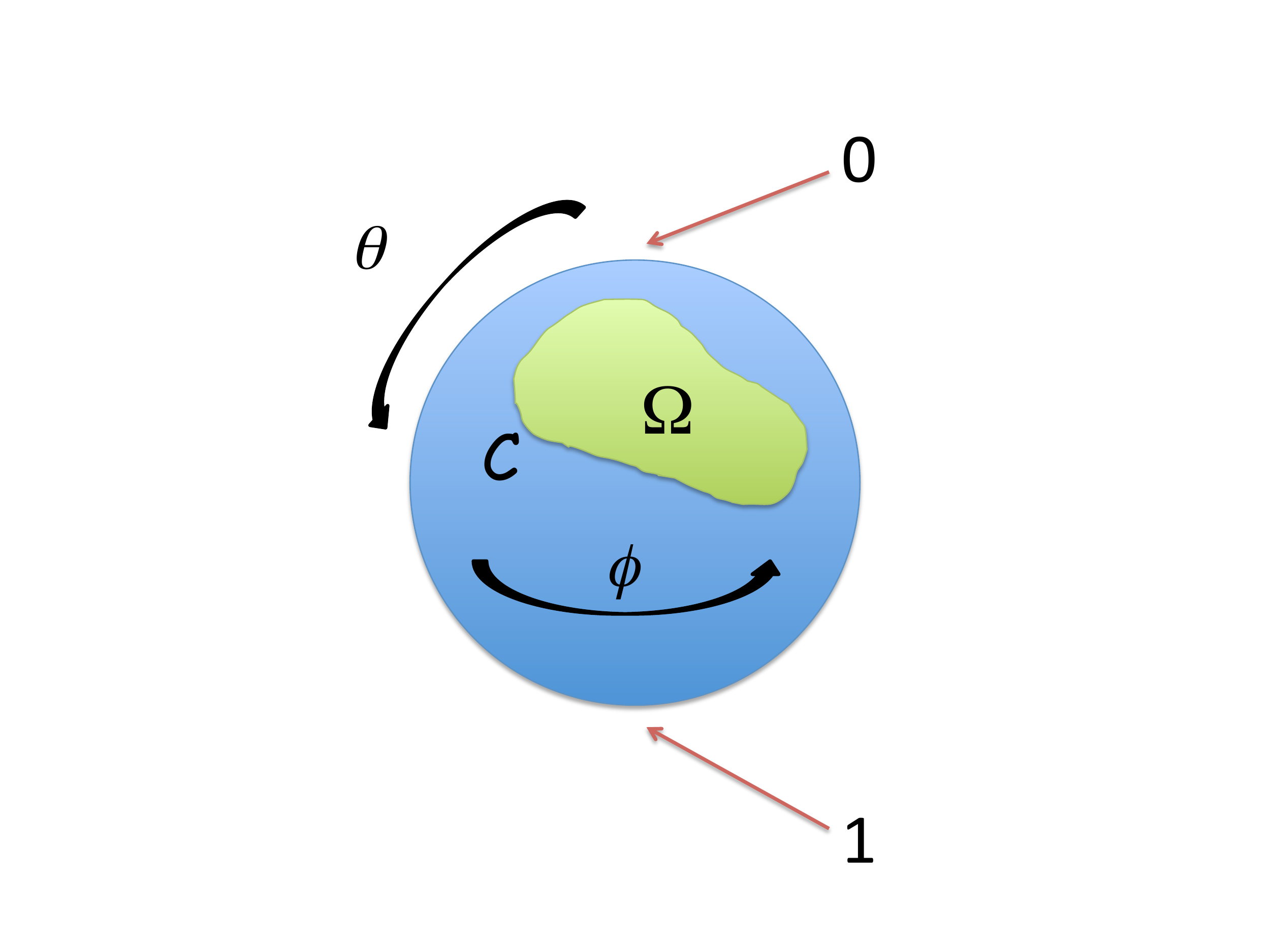}
\caption{Bloch sphere representing pure qubit states with the projection of the basis states 
$\ket{0}$ and $\ket{1}$ at the poles. The angles $\theta$ and $\phi$ determine the relative 
weight and phase shift, respectively, between $\ket{0}$ and $\ket{1}$. A loop $\mathcal{C}$ 
enclosing the solid angle $\Omega$ on the Bloch sphere is associated with the GP 
$-\frac{1}{2} \Omega$.} 
\label{fig1}
\end{figure*}

A qubit is a quantum system that can be described by two states $\ket{0}$ and $\ket{1}$. 
Physical examples of qubits are abundant: it could be the polarization of a photon, the 
spin of an electron, neutron, or proton, two isolated states of an atom or ion, two 
charge state of a superconducting island, and so on. An arbitrary qubit state $\ket{\psi}$ 
can be written as 
\begin{eqnarray}
\ket{\psi} = \cos \frac{\theta}{2} \ket{0} + e^{i\phi} \sin \frac{\theta}{2} \ket{1} . 
\end{eqnarray}
The two parameters $\theta$ and $\phi$ may be interpreted as polar angles of a unit 
sphere, called the Bloch sphere, with $\ket{0}$ and $\ket{1}$ projecting onto the north 
($\theta = 0$) and south poles ($\theta = \frac{\pi}{2}$), respectively, see Figure \ref{fig1}. 

Now, by assuming that the qubit traces out a loop $\mathcal{C} : t \in [0,\tau] \rightarrow 
(\theta_t,\phi_t)$ on the Bloch sphere, i.e., $\theta_{\tau} = \theta_0$ and 
$\phi_{\tau} = \phi_0$, we obtain  
\begin{eqnarray}
\Phi [\mathcal{C}] = - \frac{1}{2} \oint_{\mathcal{C}} (1-\cos \theta) d\phi = 
- \frac{1}{2} \Omega ,  
\end{eqnarray}
$\Omega$ being the solid angle enclosed on the Bloch sphere by $\mathcal{C}$. By 
using Stokes theorem, we obtain 
\begin{eqnarray}
\Phi [\mathcal{C}] = - \frac{1}{2} \oint_{\mathcal{S}} \sin \theta d\theta d\phi 
\end{eqnarray}
where $\mathcal{S}$ is the surface on the Bloch sphere enclosed by $\mathcal{C}$.
This is equivalent to the flux through $\mathcal{S}$ of a magnetic monopole of 
strength $-\frac{1}{2}$ sitting at the origin. 

\section*{\sffamily \Large Non-Abelian GPs} 
Following Berry's work \cite{berry84} on Abelian GP factors accomanying adiabatic 
changes, Wilczek and Zee \cite{wilczek84} pointed out that non-Abelian gauge structures 
appear when more than one state is considered. This turns out to be relevant for instance 
in adiabatic evolution where several energy eigenstates are degenerate over some region 
of parameter space. 

The key ingredient of the non-Abelian generalization is the evolution of a subspace of the 
full state space. Let $t \in [0,\tau] \rightarrow \mathcal{S}_t$ trace out a loop 
$\mathcal{C}$ (i.e., $\mathcal{S}_{\tau} = \mathcal{S}_0$) of $K$-dimensional subspaces of 
an $N$-dimensional Hilbert space. Let $\{ \ket{\psi_k (t)} \}_{k=1}^K$ span $\mathcal{S}_t$. 
Since $\mathcal{C}$ is a loop, the overlap matrix ${\bf U}_{kl} (0,{\tau}) = \bra{\psi_k (0)} \psi_l (\tau) 
\rangle$ is unitary and corresponds to the overall `phase' of the evolution.  By applying the 
same reasoning for two infinitesimally close instances $t$ and $t + \delta t$, we obtain the local 
`phase shift' $e^{i{\bf A} (t) \delta t}$ to first order in $\delta t$. Here, ${\bf A}_{kl} (t) = 
i\bra{\psi_k (t)} \dot{\psi}_l(t) \rangle$. The accumulated local phase shift along the loop is 
$\mathcal{T} e^{i\int_0^{\tau} {\bf A} (t) dt}$, where $\mathcal{T}$ stands for time ordering. 
Thus, we obtain the non-Abelian GP: 
\begin{eqnarray} 
{\bf U} [\mathcal{C}] =  {\bf U} \mathcal{T} e^{i\int_0^{\tau} {\bf A} (t) dt} , 
\end{eqnarray} 
where ${\bf U} \equiv {\bf U} (0,\tau)$. 
Note that for the special case where $K=1$, ${\bf U} [\mathcal{C}]$ reduces to the 
complex number  
\begin{eqnarray} 
U [\mathcal{C}] = e^{i\arg \bra{\psi (0)} \psi (\tau) \rangle} 
e^{-\int_0^{\tau} \bra{\psi (t)} \dot{\psi} (t) \rangle dt} = 
e^{i\left( \arg \bra{\psi (0)} \psi (\tau) \rangle + i\int_0^{\tau} \bra{\psi (t)} \dot{\psi} (t) 
\rangle dt \right)} ,  
\end{eqnarray}
which we recognize as the Abelian GP factor $e^{i\Phi [\mathcal{C}]}$ discussed above. 

The subspace need not trace out a loop to acquire a non-Abelian GP. For such an open path
$\mathcal{C}$, the overlap matrix ${\bf M}_{kl} (0,{\tau}) = \bra{\psi_k (0)} \psi_l (\tau) \rangle$ 
is no longer unitary, but admits a unique polar decomposition ${\bf M} = |{\bf M}| {\bf U}$ 
provided $|{\bf M}| > 0$. We therefore have 
\begin{eqnarray} 
{\bf M} \mathcal{T} e^{i\int_0^{\tau} {\bf A} (t) dt} = 
|{\bf M}| {\bf U} \mathcal{T} e^{i\int_0^{\tau} {\bf A (t)} dt} \equiv |{\bf M}| {\bf U} [\mathcal{C}] , 
\end{eqnarray} 
which defines the non-Abelian GP \cite{kult06} 
\begin{eqnarray}
{\bf U} [\mathcal{C}] = |{\bf M}|^{-1} {\bf M} \mathcal{T} e^{i\int_0^{\tau} {\bf A} (t) dt}
\end{eqnarray}
valid for any evolution connecting two fully overlapping $K$-dimensional subspaces 
$\mathcal{S}_0$ and $\mathcal{S}_{\tau}$ of the $N$-dimensional Hilbert space. Note 
that the existence of the inverse $|{\bf M}|^{-1}$ is a consequence of the condition 
$|{\bf M}| > 0$ that is required for a unique polar decomposition of the overlap matrix. 

The matrix-valued GPs may as well be viewed as operators: 
\begin{eqnarray}
U [\mathcal{C}] = \sum_{k,l=1}^K {\bf U}_{kl} [\mathcal{C}]  
\ket{\psi_k (\tau)} \bra{\psi_l (0)} , 
\end{eqnarray}
with initial subspace as domain and final subspace as image (these subspaces coincide if 
$\mathcal{C}$ is a loop). The matrix and operator forms of the non-Abelian GPs are 
fully equivalent. 

The non-Abelian GP transforms as ${\bf U} [\mathcal{C}] \rightarrow {\bf V} (0) {\bf U} 
[\mathcal{C}] {\bf V}^{\dagger} (0)$ under a unitary change of basis $\ket{\psi_k (t)} 
\rightarrow \sum_l \ket{\psi_l (t)} V_{lk} (t)$ along the moving subspace $\mathcal{S}_t$. 
This implies that the Wilson loop $\Tr {\bf U} [\mathcal{C}]$ is unchanged under local 
basis changes. Thus, the non-Abelian GP transforms gauge covariantly and is therefore 
a property of the path $\mathcal{C}$ in the space of subspaces \cite{kult06}. 

\subsection*{\sffamily \large Example: Geometric transformation of a qubit} 
A technical complication when calculating the non-Abelian GP is the need to evaluate 
a time-ordered product. This may be dealt with by performing a pair of loops resulting 
in non-commuting GPs, therefore demonstrating the desired non-Abelian feature, but 
along each of these loops no time-ordering is needed. Here, we illustrate this idea 
in the case of a two-dimensional subspace $\mathcal{S}$, encoding a single qubit, 
that evolves in a three-dimensional Hilbert space $\mathcal{H}$. 

Let $\mathcal{H} = \textrm{Span} \{ \ket{0}, \ket{1}, \ket{a} \}$. Consider paths of 
two-dimensional subspaces $\mathcal{C} \rightarrow \mathcal{S}_t = \textrm{Span} 
\{ \ket{\psi (t)}, \ket{\psi^{\perp} (t)} \}$ such that $\mathcal{S}_0$ is the qubit subspace. 
Thus, $\mathcal{S}_0 = \textrm{Span} \{ \ket{0}, \ket{1} \}$. 

Let us first consider the choice 
\begin{eqnarray}
\ket{\psi_1 (t)} & = & \ket{0} ,
\nonumber \\  
\ket{\psi_1^{\perp} (t)} & = & 
\cos \frac{\kappa_t}{2} \ket{1} + e^{i\eta_t} \sin \frac{\kappa_t}{2} \ket{a} . 
\end{eqnarray}
A loop $\mathcal{C}_1$ that starts and ends at $\mathcal{S}_0$ is implemented by 
choosing $\kappa_{\tau} = \kappa_0 = 0$ and $\eta_{\tau} = \eta_0$. The GP 
associated with $\mathcal{C}_1$ is 
\begin{eqnarray}
U [\mathcal{C}_1] = e^{-i \ket{1}\bra{1} 
\oint_{\mathcal{C}_1} (1-\cos \kappa) d\eta} .
\label{eq:u1}
\end{eqnarray}
The phase shift can be interpreted as the solid angle enclosed on the Bloch sphere 
associated with the subspace $\textrm{Span} \{ \ket{1},\ket{a} \}$ and parametrized 
by the spherical polar angles $\kappa,\eta$.  

Now, consider instead the choice  
\begin{eqnarray}
\ket{\psi_2 (t)} & = & -\sin \eta_t \ket{0} + \cos \eta_t \ket{1} ,  
\nonumber \\  
\ket{\psi_2^{\perp} (t)} & = & \cos \kappa_t \cos \eta_t \ket{0} + 
\cos \kappa_t \sin \eta_t \ket{1} - \sin \kappa_t \ket{a} , 
\end{eqnarray}
which span a subspace that starts and ends at $\mathcal{S}_0$ provided 
$\kappa_{\tau} = \kappa_0 = 0$ and $\eta_{\tau} = \eta_0$. The GP associated 
with this loop $\mathcal{C}_2$ is 
\begin{eqnarray}
U[\mathcal{C}_2] = 
e^{-i \sigma_y \oint_{\mathcal{C}_2} (1-\cos \kappa) d\eta} , 
\label{eq:u2}
\end{eqnarray} 
where $\sigma_y = -i \ket{0} \bra{1} + i \ket{1} \bra{0}$. Just as $U [\mathcal{C}_1]$, 
the unitary qubit transformation $U [\mathcal{C}_2]$ has a geometric interpretation in 
terms of a solid angle but with a different meaning: it is the solid angle enclosed on a 
parameter sphere that defines all possible choices of two-dimensional subspaces with 
real-valued coefficients with respect to the given reference basis $\{ \ket{0},\ket{1},\ket{a} \}$ 
of the full Hilbert space. 

For non-trivial loops $\mathcal{C}_1$ and $\mathcal{C}_2$ the resulting $U [\mathcal{C}_1]$ 
and $U [\mathcal{C}_2]$ are non-commuting thereby demonstrating the desired non-Abelian 
feature of these GPs. The above qubit GPs were discovered theoretically by Unanyan {\it et al.} 
\cite{unanyan99} in the context of atoms controlled by stimulated Raman adiabatic 
passage (STIRAP) techniques. 
 
\section*{\sffamily \Large QUANTUM COMPUTATION}
A classical computer works by manipulating a set of bits according to some algorithm 
to produce a desired result. Analogously, a quantum computer manipulates qubits by applying 
a certain combination of logical operations, quantum gates, forming a quantum circuit. GPs 
have been proposed to be useful in realizing quantum gates.  One of the main reasons for 
considering GP-based quantum gates is that they are believed to be more robust 
than traditional dynamical gates, and therefore useful in order to reach the error threshold 
\cite{knill05}, below which quantum computation with faulty gates can be performed
by means of error correction protocols \cite{shor95,steane97}.  

A key ingredient of circuit-based quantum computation is the concept of universality, which 
is the ability to perform an arbitrary unitary transformation on a certain set of qubits. It has 
been shown that universality can be achieved by applying arbitrary single-qubit operations 
and at least one entangling two-qubit gate \cite{bremner02}. An arbitrary single-qubit operation 
requires the ability to perform non-commuting gates. Geometric quantum computation (GQC) 
is the idea to use GPs to implement such a universal set of one- and two-qubit gates.  

There is basically two different methods to achieve universality by using GPs. Either, one 
realizes geometric phase shift gates with respect to different bases \cite{zhu02,zhu03a}, 
or one realizes non-commuting gates by using non-Abelian GPs for $K>1$ dimensional 
subspaces \cite{zanardi99,sjoqvist12}. Each of these methods can be implemented by 
employing adiabatic or non-adiabatic evolution. 
 
\subsection*{\sffamily \large GQC based on Abelian GPs} 
Geometric phase shift gates take the form $\ket{k} \rightarrow e^{if_k} \ket{k}$ with 
$f_k$ being the GP of the computational state $\ket{k}$. At first sight, 
such gates are not sufficient for universality since they all commute. To resolve this, one 
may instead implement GPs with respect to different bases \cite{zhu02,zhu03a}. 
The resulting phase shift gates become non-commuting and therefore potentially universal. 

To demonstrate this idea, consider a single-qubit $\ket{k}$, $k=0,1$. Assume the 
orthogonal states $\ket{+} = \cos \frac{\theta}{2} \ket{0} + e^{i\phi} \sin \frac{\theta}{2} 
\ket{1}$ and $\ket{-} = - e^{-i\phi} \sin \frac{\theta}{2} \ket{0} + \cos \frac{\theta}{2} \ket{1}$  
aquire purely GP shifts $\mp \Omega /2$ after cyclic evolution ($\Omega$ being the 
solid angle enclosed by the Bloch vector). Thus, $\ket{\pm} \rightarrow e^{\mp i\Omega/2} 
\ket{\pm}$ defining the geometric gate transformation \cite{zhu02,zhu03a} 
\begin{eqnarray}
U (\Omega,{\bf n}) = e^{-i\Omega/2} \ket{+} \bra{+} + e^{i\Omega/2} \ket{-} \bra{-} = 
e^{-i \frac{1}{2} \Omega {\bf n} \cdot \boldsymbol{\sigma}} , 
\end{eqnarray} 
where $\boldsymbol{\sigma} = (\sigma_x,\sigma_y,\sigma_z)$ are the Pauli operators 
and ${\bf n} = (\sin \theta \cos \phi , \sin \theta \sin \phi , \cos \theta )$. Provided 
the solid angle and ${\bf n}$ can be varied independently, $U (\Omega,{\bf n})$ is an 
arbitrary SU(2) transformation, i.e., a universal one-qubit gate. 

The dynamical phase contributions of the $\ket{\pm}$ states can be eliminated either 
by employing refocusing technique \cite{jones99,ekert00}, rotating driving fields with 
fine-tuned parameters \cite{zhu02,zhu03a,zhu05}, or by driving the qubit along geodesics 
on the Bloch sphere by using composite pulses \cite{tian04}. These techniques result 
in non-commuting gates solely dependent on the GPs of the cyclic states.  

\subsubsection*{\sffamily \normalsize Physical implementation: NMR qubits} 
Quantum computation based on nuclear magnetic resonance (NMR) technique manipulates 
qubits encoded in nuclear spin-$\frac{1}{2}$. Such a qubit evolves due to the Zeeman 
interaction $\mu {\bf I} \cdot {\bf B}$, $\mu$ being the magnetic moment of the nuclei, 
${\bf I}$ the spin of the nuclei, and ${\bf B}$ the applied magnetic field. 

The nuclear spin states $\ket{\uparrow ; {\bf n}}$ and $\ket{\downarrow ; {\bf n}}$, ${\bf n}$ 
being the quantization axis, acquire GPs $-\Omega /2$ and $+\Omega /2$, respectively, 
by adiabatically turning the magnetic field around a loop $\mathcal{C}$ that starts and 
ends along ${\bf n}$ and encloses the solid angle $\Omega$. The dynamical phases are 
eliminated by refocusing, whereby the spin is taken around $\mathcal{C}$ twice, the 
second time exactly retracing the first evolution but in opposite dirrection, and performing 
fast spin flips ($\pi$ transformations) immediately before and after the second loop. Thus, 
the spin undergoes the evolution $\mathcal{C} \rightarrow \pi \rightarrow \mathcal{C}^{-1} 
\rightarrow \pi$. While the GPs of the qubit spin states $\ket{\uparrow ; {\bf n}}$ and 
$\ket{\downarrow ; {\bf n}}$ add up after performing the spin echo sequence, the 
dynamical phases exactly cancel leaving a purely geometric phase shift gate 
\begin{eqnarray}
U (2\Omega,{\bf n}) = e^{-i \Omega {\bf n} \cdot \boldsymbol{\sigma}} ,  
\end{eqnarray}
which is a universal one-qubit operation as noted above. A key point of the refocusing 
technique is that it makes the scheme insensitive to spatial variations of the magnetic 
field strength so that all spin qubits in the sample undergo the same purely geometric 
transformation, determined by the same solid angle $\Omega$ swept by the direction 
of the slowly varying magnetic field \cite{jones99}. 

Next, a conditional phase shift gate acting on two nuclear spin qubits can be realized by 
utilizing the standard NMR uniaxial spin-spin interaction $J I_{z;a} \otimes I_{z;b}$ in addition 
to the Zeeman term $\left( \mu_a {\bf I}_a + \mu_b {\bf I}_b \right) \cdot {\bf B}$ of the two 
spins $a$ and $b$. The interaction term effectively shifts the $z$-component of the magnetic 
field by $J/(\mu_a+\mu_b)$ or $-J/(\mu_a+\mu_b)$ depending on whether the spins are 
parallel or anti-parallel. By choosing the initial quantization axis of the two spins along the 
$z$-direction, the sequence $\mathcal{C} \rightarrow \pi_a \rightarrow \mathcal{C}^{-1} 
\rightarrow \pi_b \rightarrow \mathcal{C} \rightarrow \pi_a \rightarrow \mathcal{C}^{-1} 
\rightarrow \pi_b$, where $\pi_a$ and $\pi_b$ are spin flips applied selectively to spins 
$a$ and $b$, respectively, and  $\mathcal{C}$ is an adiabatic loop of the magnetic field, 
results in the conditional phase shift gate \cite{jones99}
\begin{eqnarray}
U (\Delta \gamma) & = & e^{2i\Delta \gamma} \big( \ket{\uparrow_a \uparrow_b ; z} 
\bra{\uparrow_a \uparrow_b ; z} + \ket{\downarrow_a \downarrow_b ; z} 
\bra{\downarrow_a \downarrow_b ; z}  \big) 
\nonumber \\ 
 & & + e^{-2i\Delta \gamma} \big( \ket{\uparrow_a \downarrow_b ; z} 
\bra{\uparrow_a \downarrow_b ; z} + \ket{\downarrow_a \uparrow_b ; z} 
\bra{\downarrow_a \uparrow_b ; z}  \big) .  
\label{eq:cpsg}
\end{eqnarray} 
Here, 
\begin{eqnarray} 
\Delta \gamma = \gamma_+ - \gamma_- , 
\end{eqnarray}
where $\gamma_{\pm} = \Omega_{\pm} /2$ with $\Omega_{\pm}$ being the solid angles 
enclosed by the effective magnetic fields $\big( B_x,B_y,B_z \pm J/(\mu_a + \mu_b) \big)$.  

Note that $U (\Delta \gamma)$ is entangling as it cannot be written as a product of 
unitary operators acting locally on each nuclear spin. Thus, $U (2\Omega,{\bf n})$ and 
$U (\Delta \gamma)$ form a universal set of one- and two-qubit gates that can be used 
to perform any quantum computation by purely geometric means. 

\subsubsection*{\sffamily \normalsize Experiments} 
The conditional geometric phase shift gate $U (\Delta \gamma)$ has been demonstrated 
in NMR by Jones {\it et al.} \cite{jones99}. The two qubits were taken as the weakly 
interacting nuclear spins. The desired geometric gate was realized by adiabatically 
sweeping the magnetic field around a loop $\mathcal{C}$ starting and ending along the 
$z$ axis. 

Another class of experiments related to GQC is based on the idea that the dynamical phase 
can in some cases be proportional to the geometric phase \cite{zhu03b}. This avoids the 
need to eliminate the dynamical phase. Since such gates  have non-zero dynamical phase, 
they are not geometric gates in a strict sense; nevertheless they share the potential 
robustness of standard geometric gates. The term `unconventional GQC' \cite{zhu03b} 
has been coined for this type of gates. 

In \cite{leibfried02}, a conditional $\pi$-phase two-qubit unconventional geometric gate 
combined with single-qubit rotations to produce a Bell state with 97\% fidelity of two trapped 
ions has been demonstrated. A universal set of one- and two-qubit unconventional geometric 
gates with average fidelities 97-99\% (one-qubit gates) and 93\% (two-qubit gate) has been 
demonstrated in NMR \cite{du06}. These high fidelities indicate the potential usefulness of 
Abelian geometric phases for robust quantum computation.  
 
\subsection*{\sffamily \large Holonomic quantum computation: GQC based on non-Abelian GPs} 
Non-Abelian GPs are matrix-valued and can therefore be non-commuting. By finding methods 
to eliminate the dynamical phases, all-geometric universal quantum computation can thus be 
implemented by using non-Abelian GPs. 

In the standard scheme of non-Abelian GQC \cite{zanardi99,pachos99,pachos01}, the 
dynamical phase is eliminated by utilizing adiabatic evolution of energetically degenerate 
subspaces. In this way, matrix-valued Wilczek-Zee phases \cite{wilczek84} are implemented 
that can be used to realize a universal set of quantum gates \cite{duan01,recati02,faoro03,solinas03}. 
A difficulty in realizing these gates is the long run time associated with adiabatic evolution. 
In other words, adiabatic non-Abelian geometric gates operate slowly compared to the dynamical 
time scale, which make them vulnerable to open system effects and parameter fluctuations that 
may lead to loss of coherence. 

To overcome the problem with the long run-time, GQC based on non-adiabatic non-Abelian 
GPs has been proposed \cite{sjoqvist12} and further developed 
\cite{azimi14,gungordu14,malinovsky14}. The non-adiabatic scheme can be performed 
at high speed and involves less parameters to control experimentally than in the standard 
adiabatic scheme. 

In the following, we describe the basic idea of adiabatic and non-adiabatic non-Abelian GQC.  
We describe the schemes in terms of trapped atoms or ions, which provide a versatile tool to 
perform quantum information processing by geometric means. Qubits encoded in the energy 
levels of the atoms or ions can be manipulated by external fields and the trapping potential to 
realize universal non-adiabatic non-Abelian GPs. 

\subsubsection*{\sffamily \normalsize Non-Abelian GQC: Adiabatic case} 
Adiabatic non-Abelian GQC can be implemented in an atomic or ionic system by controlling 
transitions between four energy levels by three independent laser pulses forming a tripod 
configuration. This scheme, first proposed for trapped ions or atoms \cite{duan01,recati02} 
(see also \cite{pachos02,moller07}), and later for superconducting qubits \cite{faoro03} and 
quantum dots \cite{solinas03}, has become the standard one to perform GQC by means 
of adiabatic evolution. Other schemes not based on the tripod configuration can be found in 
\cite{niskanen03,karle03,tanimura04,pirkkalainen10}.  

The tripod shown in Figure \ref{fig2}, consists of four atomic or ionic energy levels 
$\ket{0},\ket{1},\ket{a}$ and $\ket{e}$ of the `bare' Hamiltonian $H_0 = -\hbar 
\left( f_{e0} \ket{0} \bra{0} +f_{e1} \ket{1} \bra{1} + f_{ea} \ket{a} \bra{a} \right)$ 
controlled by suitable external fields (by putting the `bare' energy of the excited 
state to zero). The states $\ket{0},\ket{1},\ket{a}$ couple to the `excited' state 
$\ket{e}$ by applying three oscillating electric field pulses ${\bf E}_j (t) = 
\boldsymbol{\epsilon}_j g_j(t/T) \cos (f_j t)$, $j=0,1,a$, where the $g_j$'s describe slowly 
varying shapes and relative phases of the pulses, and the $\boldsymbol{\epsilon}_j$'s 
describe the polarization of the laser beams. By tuning the oscillation 
frequencies $f_j$ so that detunings satisfy $f_j - f_{ej} \equiv \Delta$, the Hamiltonian 
in the interaction picture reads 
\begin{eqnarray}
H_I^{(1)} (t) & = & -\hbar \Delta \ket{e} \bra{e} + 
\hbar (\omega_0 (t/T) \ket{e}\bra{0} + \omega_1 (t/T) \ket{e}\bra{1} + 
\omega_a (t/T) \ket{e}\bra{a}) + {\textrm{h.c.}} , 
\end{eqnarray}
where $\omega_j = \bra{e} \boldsymbol{\mu} \cdot \boldsymbol{\epsilon}_j \ket{j} 
g_j /(2\hbar)$, $\boldsymbol{\mu}$ being the electric dipole moment operator, 
and we have neglected rapidly oscillating counter-rotating terms $e^{\pm2if_{ej} t}$ 
terms (rotating wave approximation (RWA)). 

\begin{figure*}[ht]
\centering
\includegraphics[width=0.9\textwidth]{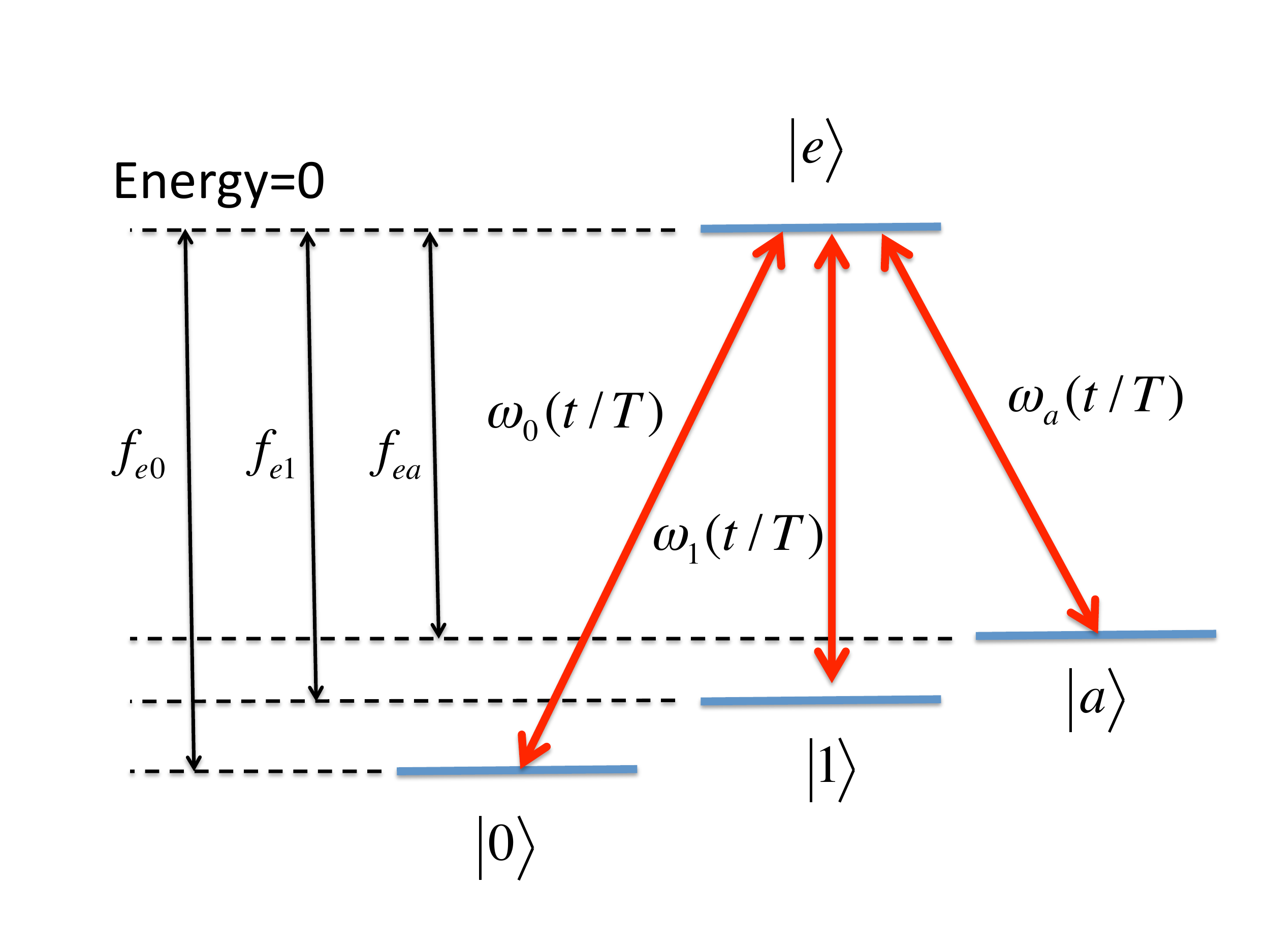}
\caption{Tripod setting consisting of three energy levels $\ket{0},\ket{1}$, and $\ket{a}$ 
coupled to an excited state $\ket{e}$ at zero energy by laser pulses with the same detuning. 
Adiabatic non-Abelian geometric gates can be implemented by slowly varying the 
complex-valued laser parameters $\omega_j$, $j=0,1,a$, around a loop in parameter 
space. The parameters start and end so that the two-dimensional dark subspace coincide 
with the qubit subspace $\textrm{Span} \{ \ket{0},\ket{1} \}$.} 
\label{fig2}
\end{figure*}
 
The interaction Hamiltonian $H_I^{(1)} (t)$ has two degenerate dark energy eigenstates: 
\begin{eqnarray}
\ket{D_1;\theta,\phi} & = & -\sin \varphi e^{i(S_3 - S_1)} \ket{0} + \cos \varphi 
e^{i(S_3 -S_2)} \ket{1} , 
\nonumber \\  
\ket{D_2;\theta,\phi} & = & \cos \vartheta \left( \cos \varphi e^{i(S_3 - S_1)} \ket{0} + 
\sin \varphi e^{i(S_3 - S_2)} \ket{1} \right) - \sin \vartheta \ket{a} , 
\end{eqnarray}
where $\omega_0 = \omega \sin \vartheta \cos \varphi e^{iS_1}$, $\omega_1 = 
\omega \sin \vartheta \sin \varphi e^{iS_2}$, and $\omega_a = \omega \cos \vartheta 
e^{iS_3}$. These dark eigenstates coincide with the realization of the non-commuting 
one-qubit operations $U[\mathcal{C}_1]$ and $U[\mathcal{C}_2]$ in Eqs. (\ref{eq:u1}) 
and (\ref{eq:u2}) by identifying the adiabatic parameters $(\vartheta , \varphi , S_3 - S_1,  
S_3 - S_2)$ with $(\kappa /2,\pi /2, 0 ,-\eta)$ and $(\kappa ,\eta, 0 ,0)$, respectively. 
These two gates are realized by varying the parameters $\kappa,\eta$ in the  
$T \rightarrow \infty$ adiabatic limit. 

The above $U[\mathcal{C}_1]$ and $U[\mathcal{C}_2]$ form a universal set together 
with any entangling geometric two-qubit gate. For trapped ions, such a gate can be performed 
by utilizing the S{\o}rensen-M{\o}lmer setting \cite{sorensen99}, resulting in the 
Hamiltonian \cite{duan01}
\begin{eqnarray}
H_I^{(2)} (t) & = & \omega \left( - \sin \frac{\kappa}{2} e^{i\eta} \ket{ee} \bra{11} + 
\cos \frac{\kappa}{2} \ket{ee} \bra{aa} + \textrm{h.c.} \right) , 
\end{eqnarray} 
which has a single dark state $\ket{D;\kappa,\eta} = \cos \frac{\kappa}{2} \ket{11} + 
e^{i\eta} \sin \frac{\kappa}{2}\ket{aa}$ that picks up the GP 
\begin{eqnarray} 
U[\mathcal{C}_3] = e^{-i \ket{11} \bra{11} \oint (1-\cos \kappa) d\eta} 
\end{eqnarray} 
by adiabatically changing $\kappa$ and $\eta$ around a loop in prameter space. 
$U[\mathcal{C}_1],U[\mathcal{C}_2]$, and $U[\mathcal{C}_3]$ form a universal set 
of geometric quantum gates that can be used to build any quantum computation 
by purely geometric means. 

\subsubsection*{\sffamily \normalsize Non-Abelian GQC: Non-adiabatic case} 
Non-adiabatic non-Abelian GQC \cite{sjoqvist12} in an atomic or ionic system requires 
control of three energy levels by suitable external laser fields forming a $\Lambda$-system. 
By following the same procedure as above, we obtain the Hamiltonian 
\begin{eqnarray}
H_I^{(1)} (t) & = & \omega_0(t) \ket{e}\bra{0} + \omega_1(t) \ket{e}\bra{1} + 
{\textrm{h.c.}}  
\end{eqnarray}
by employing the RWA in the interaction picture and by requiring zero detuning. A key 
difference compared to the adiabatic scheme is that the complex-valued Rabi frequencies 
$\omega_0 (t),\omega_1 (t)$ are allowed to be varied at any speed.  

\begin{figure*}[ht]
\centering
\includegraphics[width=0.9\textwidth]{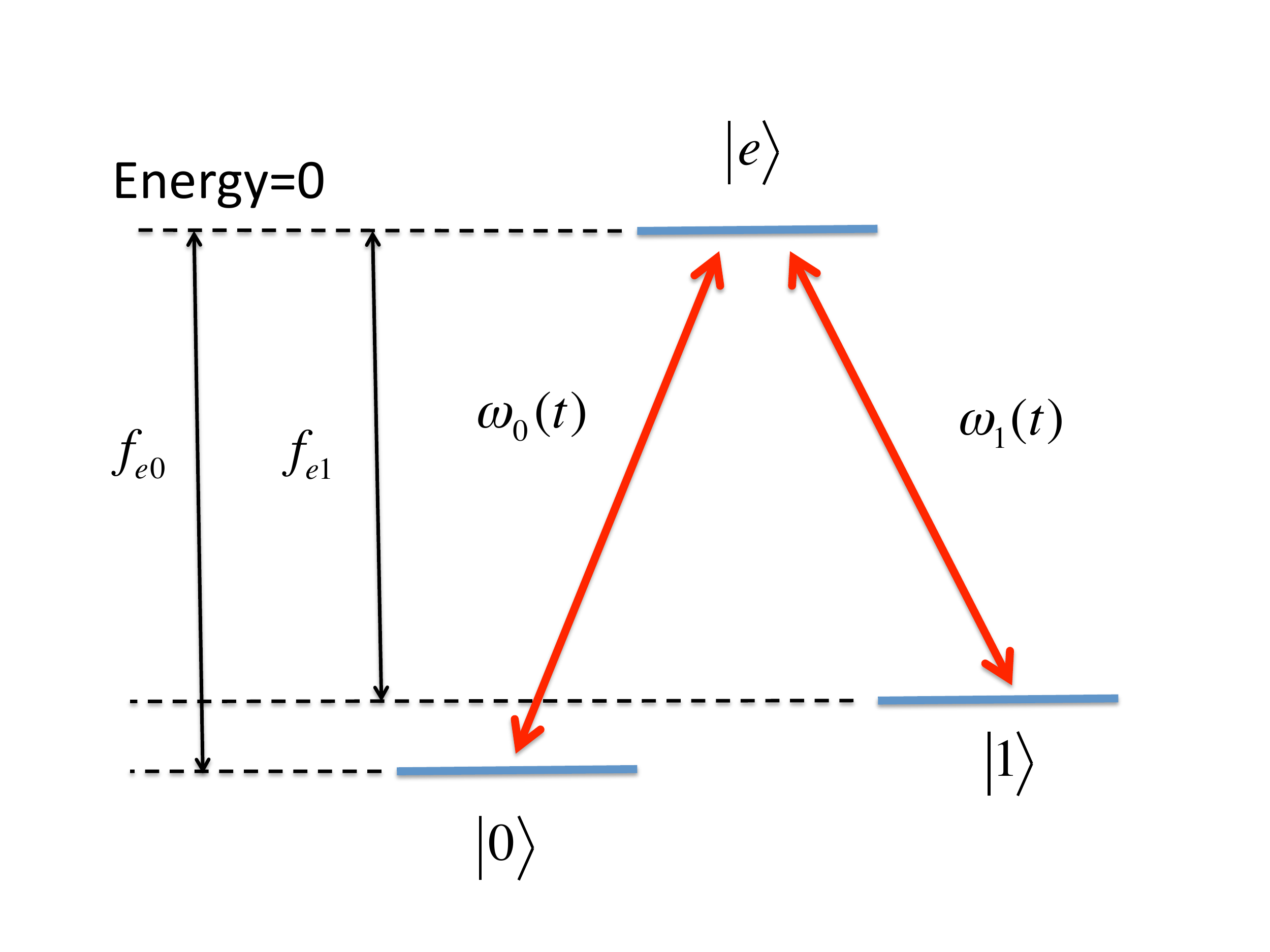}
\caption{$\Lambda$ setting consisting of two energy levels $\ket{0}$ and $\ket{1}$ 
coupled to an excited state $\ket{e}$ at zero energy by a pair of zero-detuned laser pulses. 
Non-adiabatic non-Abelian geometric gates acting on a qubit encoded in the subspace 
$\textrm{Span} \{ \ket{0},\ket{1} \}$ can be implemented by varying the complex-valued 
laser parameters $\omega_j$, $j=0,1,a$, but keeping $\omega_0 (t)/\omega_1 (t)$ 
constant over each pulse pair chosen to satisfy the $\pi$ pulse criterion $\int_0^{\tau} 
\sqrt{|\omega_0 (t)|^2 + |\omega_1 (t)|^2} dt = \pi$.} 
\label{fig2}
\end{figure*}
 
A non-Abelian geometric one-qubit gate $U[\mathcal{C}]$ acting on the qubit subspace 
is implemented by choosing laser field  pulses of duration $\tau$ such that $\omega_0 (t) 
/ \omega_1 (t) \equiv -\tan (\theta /2) e^{i\phi}$ is time independent and satisfy the 
criterion $\int_0^{\tau} \sqrt{|\omega_0 (t)|^2 + |\omega_1 (t)|^2} dt = \pi$. Here, the 
latter condition assures that the qubit subspace $\textrm{Span} \{ \ket{0},\ket{1} \}$ 
undergoes a cyclic evolution defining a loop 
$\mathcal{C}$ in the space of two-dimensional subspaces of the three-dimensional 
Hilbert space $\textrm{Span} \{ \ket{0},\ket{1},\ket{e} \}$; the former guarantees that 
the dynamical phases vanish for the full pulse duration, which implies that the gate is 
fully determined by $\mathcal{C}$. Explicitly, 
\begin{eqnarray}
U[\mathcal{C}] = {\bf n} \cdot \boldsymbol{\sigma}, 
\end{eqnarray}
where $\boldsymbol{\sigma}$ are the Pauli operators acting on the qubit subspace. 
This unitary corresponds to a $180^{\circ}$ rotation of the qubit around the `direction' 
${\bf n} = (\sin \theta \cos \phi ,\sin \theta \sin \phi ,\cos \theta)$ defined by relative 
phase and amplitude of the pair of laser fields. Performing sequentially two such gates 
with different laser parameters defining directions ${\bf n}$ and ${\bf n}'$ yields 
\begin{eqnarray}
U(\mathcal{C}') U[\mathcal{C}] = {\bf n}' \cdot {\bf n} + i\boldsymbol{\sigma} 
\cdot ({\bf n}' \times {\bf n} ) ,
\end{eqnarray} 
which is an arbitrary SU(2) operation, i.e., a universal one-qubit gate. Geometrically, 
the gate can be visualized as a rotation of the qubit by an angle $-2\arccos \left( 
{\bf n}' \cdot {\bf n} \right)$ around the direction ${\bf n}' \times {\bf n}$. 

Similar to the adiabatic realization, the universal set is completed by adding a geometric 
two-qubit gate in the S{\o}rensen-M{\o}lmer setting \cite{sorensen99}. The differences 
are that each ion needs to exhibit only an internal three-level structure $\ket{0}, \ket{1}$ 
and $\ket{e}$, and the amplitude ratio $\tan (\theta /2)$ as well as the phase shift $\phi$ 
of two laser beams should be kept constant during each pulse pair. The resulting 
Hamiltonian acting on the computational subspace $\{ \ket{00},\ket{01},\ket{10},\ket{11} \}$
reads 
\begin{eqnarray}
H_I^{(2)} =  \omega (t) \left( \sin \frac{\theta}{2} e^{i\phi /2} \ket{ee} \bra{00} -
\cos \frac{\theta}{2} e^{-i\phi /2} \ket{ee} \bra{11} + {\textrm{h.c.}} \right) . 
\end{eqnarray}
The $\pi$ pulse criterion $\int_0^{\tau} \omega (t) dt = \pi$ results in the geometric two-qubit 
gate
\begin{eqnarray}
U [\mathcal{C}] & = & \cos \theta \ket{00} \bra{00} + \sin \theta e^{-i\phi} \ket{00} \bra{11} 
 \nonumber \\ 
 & & + \sin \theta e^{i\phi} \ket{11} \bra{00} - \cos \theta \ket{11} \bra{11} 
 \nonumber \\
 & & + \ket{01} \bra{01} + \ket{10} \bra{10} . 
\label{eq:2gate}
\end{eqnarray}
The path $\mathcal{C}$, being characterized by the unit vector ${\bf n} = ( \sin \theta \cos \phi , 
\sin \theta \sin \phi , \cos \theta )$ in $\mathbb{R}^3$, is traversed in the three dimensional 
subspace spanned by $\{ \ket{00},\ket{11}, \ket{ee} \}$ of the internal degrees of freedom of 
the ions. Note that $U [\mathcal{C}]$ is entangling as it cannot be written as a product of unitary 
operators acting locally on each ion qubit. It completes the universal set of non-adiabatic 
non-Abelian geometric gates that can be used to perform fast quantum computation by 
purely geometric means. 

\subsubsection*{\sffamily \normalsize Experiments} 
The scheme for non-adiabatic GQC proposed in \cite{sjoqvist12} has been realized in several 
recent experiments \cite{abdumalikov13,feng13,arroyo-camejo14,zu14}. Non-commuting 
operations in a superconducting transmon one-qubit system have been demonstrated 
\cite{abdumalikov13}. A universal set of one- and two-qubit non-Abelian geometric gates 
has been implemented \cite{feng13} in a liquid state NMR quantum information processor 
using a three-qubit variant of \cite{sjoqvist12} akin to \cite{xu12}. 

To achieve GQC at room temperature in a naturally scalable system, spin qubits associated 
with a nitrogen-vacancy color centre in diamond have been used \cite{arroyo-camejo14,zu14}. 
While \cite{arroyo-camejo14} was limited to one-qubit operations, the full universal set of quantum 
gates, including a universal CNOT gate, which is CNOT$\ket{x}\otimes \ket{y} = \ket{x} \otimes 
\ket{x \oplus y}$ ($\oplus$ is addition mod 2 and $x,y$ take values $0$ or $1$), has been 
implemented \cite{zu14}. This CNOT gate applied to an initial product state has been shown 
to yield a concurrence \cite{wootters98} of 0.85, which unambiguously confirmed its entangling 
nature.

A recent experiment \cite{toyoda13} has demonstrated GQC using adiabatic evolution in 
a tripod configuration, where three ground state levels couple to an excited stated by use 
of three resonant laser fields. This experiment realized a universal set of one-qubit gates 
based on the adiabatic scheme proposed in \cite{duan01} resulting from adiabatic transport 
of this dark energy eigensubspace for a single trapped ion. 

\subsection*{\sffamily \large Robustness of GQC} 
A key motivation for GQC is the potential robustness of GPs to errors, such as decoherence 
and parameter noise \cite{pachos01}. To examine the validity of this conjecture, several 
studies have been carried out, testing the behavior of different forms of GQC to different 
kinds of errors. We summarize a few of these as follows: 

\begin{itemize}

\item To address the question whether geometric gates are more robust than dynamical 
gates, Zhu and Zanardi \cite{zhu05} have designed a scheme where the two types of gates 
can be continuously changed into each other. The basic idea is to implement a one-qubit 
phase shift gates $U(\gamma) = e^{i\gamma} \ket{+} \bra{+} +  e^{-i\gamma} \ket{-} 
\bra{-} $ by exposing the qubit to a rotating magnetic field. By suitably changing the 
frequency and opening angle of the magnetic field, $\gamma$ can be varied from a purely 
dynamical phase $\delta$ to a purely geometric phase $\Phi [\mathcal{C}]$. The optimal 
fidelity under parameter noise is obtained for a purely geometric gate. Although this 
result provides evidence for the advantage of GQC with regard to resilience to dephasing 
errors in this setting, there are other settings where the geometric approach seems to 
have no particular advantage compared to a dynamical approach when considering certain 
decoherence \cite{nazir02} and parameter noise \cite{blais03} models. 

\item The conditional two-qubit geometric phase gate in Eq. (\ref{eq:cpsg}) exhibits optimal 
parameter values, where $\Delta \gamma$ is resilient to errors in the amplitude of the 
RF field \cite{ekert00}. Similar optimal working points have been found at certain run-times  
in finite time realization of non-Abelian GQC under influence of an oscillator bath 
\cite{trullo06,florio06} and parameter noise \cite{lupo07}. 

\item The robustness of adiabatic and non-adiabatic realizations of non-Abelian GQC 
to parameter errors, decay, and dephasing have been examined \cite{johansson12b}. 
The adiabatic gates are robust to decay and mean detuning error in the large run-time 
limit, but they are highly sensitive to dephasing and relative detuning error in this limit. 
The non-adiabatic gates, on the other hand, become resilient to all these imperfections 
by employing pulses that are sufficiently short. However, there is a limit for how short 
the pulses can be before RWA breaks down. In fact, the experiment in \cite{abdumalikov13} 
did use pulse lengths close to this limit. Thus, one can predict that further speed-up 
in this setup would lead to highly unstable gates \cite{spiegelberg13}. 

\item Corrections to the GP of a spin-$\frac{1}{2}$ in the adiabatic $T\rightarrow \infty$ 
limit have been found to decay as $1/T$, while the corrections to the 
dynamical phase grows as $T$, by assuming a physically reasonable parameter noise 
model \cite{dechiara03}. This effect has been studied experimentally for microwave-driven  
superconducting qubits \cite{leek07,berger13} and trapped polarized ultracold neutrons 
\cite{filipp09}. 

\item Robustness can be improved by combining GQC with other error resilient schemes. 
Adiabatic non-Abelian GQC have been combined \cite{wu05} (see also 
\cite{oreshkov09a,renes13,zheng14,zheng15}) with the theory of decoherence free subspaces 
(DFSs) \cite{zanardi97,lidar98} and noiseless subsystems \cite{knill00}. This idea has been 
generalized to the non-adiabatic case \cite{xu12,zhang14a}. It has further been proved 
that GQC is scalable under a reasonable noise model by combining it with fault tolerant 
quantum error correction \cite{oreshkov09b,oreshkov09c}. All these approaches are 
experimentally challenging since they requires 3- or 4-body terms in the underlying 
Hamiltonian, while only 2-body terms occur naturally in nature. This problem may be 
addressed by using perturabtive gadgets \cite{jordan08} to simulate many-body interactions. 
An alternative solution is the recent proposal \cite{xu14a} to combine the Zhu-Wang approach 
to Abelian GQC \cite{zhu02,zhu03a} and the theory of DFSs, which removes the need for many-body 
terms but retains the universality of the geometric scheme. Finally, non-adiabatic GQC has 
been combined with dynamical decoupling \cite{xu14b}, i.e., the idea to average out the 
effect of noise by fine-tuned spin flipping \cite{viola99}. 

\end{itemize} 

\section*{\sffamily \Large MIXED STATES} 
The concept of pure quantum states is an idealization. In experiments, mixtures of 
pure states arise naturally due to inevitable imperfections in the preparation procedure and 
open system effects during evolution. Therefore, a more realistic description of experiments 
requires the notion of mixed quantum states. 

The mathematical representation of mixed states is given by density operators, which are 
linear operators $\rho$ satisfying $\rho \geq 0$ and $\Tr \rho =1$. For a mixture of pure 
states $\{ \psi_j \}$ occuring with relative frequencies $\{ p_j \}$, one can write $\rho = 
\sum_j p_j \ket{\psi_j} \bra{\psi_j}$, which explicitly entails the relation between the mixture 
$\{ p_j,\psi_j \}$ and its mathematical representation $\rho$. The density operator contains 
all empirically available information in the sense that it can be used to fully predict 
probabilities of outcomes in any conceivable experiment on a given quantum system. 

Now, to examine the robustness of GQC under influence of open system effects as well as 
imprecise preparation, we need to understand the meaning of GP associated with evolution of 
mixed quantum states. The basic question is, in other words, how to associate a physically 
meaningful GP to a path $t \in [0,\tau] \rightarrow \rho (t)$ of mixed states. For closed 
system evolution, such a path is a continuous one-parameter family of unitary transformations 
of the input density operator $\rho (0)$, i.e., $\rho (t) = U(t) \rho (0) U^{\dagger} (t)$; for an 
open system the evolution takes the form $\rho (t) = e^{\int_0^t \mathcal{L} (t') dt'} \rho (0)$, 
where $\mathcal{L}$ is a superoperator. For instance, in the Markovian limit case, $\mathcal{L}$ 
takes the Lindblad form \cite{lindblad73} $\mathcal{L} (t) \rho = -(i/\hbar) [H(t), \rho] + 
\sum_k \left( L_k \rho L_k^{\dagger} - (1/2) \{ \rho , L_k^{\dagger} L_k \} \right)$, where 
the Lindblad operators $L_k$ are arbitrary linear operators and $[\cdot , \cdot]$ 
($\{ \cdot , \cdot\}$) is the commutator (anti-commutator).  

There are two main forms of GP for mixed states: 
\begin{itemize}

\item[(i)] GP based on interferometry \cite{sjoqvist00a}. Here, the idea is to start from a 
Mach-Zehnder setup shown in Figure \ref{fig4}, where an incoming beam carrying an 
internal state $\rho$ (spin, say) is split by a 50-50 beam-splitter, the beam-pair brought 
together by two mirrors to finally interfer at a second 50-50 beam-splitter. The internal 
state $\rho$ is assumed to be unaffected by the beam-splitters and mirrors. By exposing 
one of the beams to a unitary $U$, acting on the internal state $\rho$, and the other 
beam to a U(1) phase shift $e^{i\chi}$, we obtain the output intensity 
\begin{eqnarray} 
I \propto 1 + |\Tr (\rho U)| \cos [\chi - \arg \Tr (\rho U)] 
\end{eqnarray} 
in one of the output beams. Thus, the interference oscillations produced by varying 
$\chi$ is characterized by the phase shift $\arg \Tr (\rho U)$ and visibility $|\Tr (\rho U)|$. 
The shift $\arg \Tr (\rho U)$ is the Pancharatnam relative phase \cite{pancharatnam56,sjoqvist00a} 
acquired by the internal state $\rho$ exposed to a unitary $U$. 

\begin{figure*}[ht]
\centering
\includegraphics[width=0.75\textwidth]{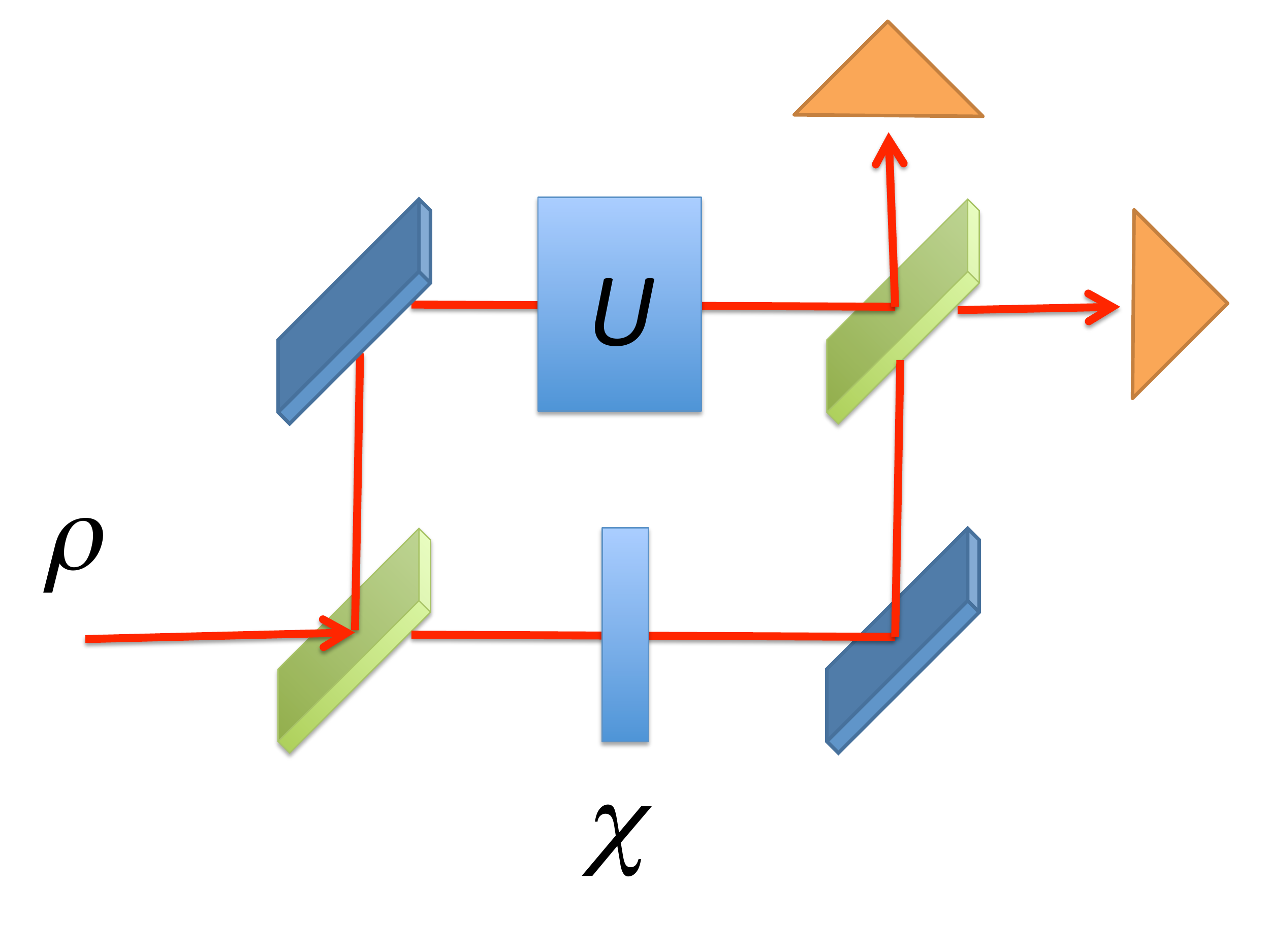}
\caption{Mach-Zehnder interferometer setup to measure the Pancharatnam relative phase 
$\arg \Tr (\rho U)$ acquired by an internal state $\rho$ exposed to a unitary $U$ in 
the upper beam. The phase is measured as a shift in the interference oscillations in 
one of the output beams obtained by varying the U(1) shift $\chi$ in the lower beam.} 
\label{fig4}
\end{figure*}

To see how this setting can be used to define a geometric phase associated with the unitary 
path $\mathcal{C}: t\in [0,\tau] \rightarrow U(t) \rho U^{\dagger} (t)$, we use the spectral 
decomposition of the density operator, i.e., $\rho = \sum p_n \ket{n} \bra{n}$ with $p_n$ 
being time-independent probabilities of the orthonormal eigenstates $\ket{n}$ of $\rho$. 
We obtain 
\begin{eqnarray}
\Tr [\rho U(\tau)] = 
\sum_{n=1}^N p_n \bra{n} U(\tau) \ket{n} = \sum_{n=1}^N p_n |\bra{n} U(\tau) \ket{n}| e^{if_n} ,  
\end{eqnarray}  
$f_n$ being the global phase acquired by the eigenstate $\ket{n}$ under the evolution 
and $N$ being the dimension of the system's Hilbert space. Each such global phase contain 
a geometric part $\arg \bra{n} U(\tau) \ket{n} + i\int_0^{\tau} \bra{n} U^{\dagger} (t) \dot{U} (t) 
\ket{n} dt$ and a dynamical part $-i\int_0^{\tau} \bra{n} U^{\dagger} (t) \dot{U} (t) \ket{n} dt$. 
By demanding parallel transport of each such eigenstate, we obtain the geometric phase shift \cite{sjoqvist00a} 
\begin{eqnarray}
\Phi [\mathcal{C}] = \arg \sum_{n=1}^N p_n \bra{n} U(\tau) \ket{n} e^{i\int_0^{\tau} 
\bra{n} U^{\dagger} (t) \dot{U} (t) \ket{n} dt} , 
\label{eq:msgp_unitary}
\end{eqnarray}
thus the GP factor $e^{i\Phi [\mathcal{C}]}$ is the weighted sum of GP factors of the 
eigenstates of $\rho$ with weights $p_n \left| \bra{n} U(\tau) \ket{n} \right|$. This GP 
is experimentally accessible since there are exactly $N$ independent phase factors 
$\{ e^{i\alpha_n (t)} \}_{n=1}^N$ that are not determined by the evolution, as can be seen by 
noting that the $U(t) \rho U^{\dagger} (t)$ is unchanged under the transformation 
$U(t) \rightarrow U(t) \sum_{n=1}^N e^{i\alpha_n (t)} \ket{n} \bra{n}$. These phase factors 
can be used to realize parallel transport of each eigenstate of the density operator. 

Provided $\rho$ is non-degenerate, $\Phi [\mathcal{C}]$ in Eq. (\ref{eq:msgp_unitary}) 
is a unique concept of GP. For a degenerate eigenvalue, however,  the corresponding 
eigenstates are not uniquely given and therefore the above definition does not provide 
a unique GP. The generalization to degenerate density operators has therefore been carried 
out \cite{singh03}, based on the idea that the degenerate subspace(s) can be associated 
non-Abelian GPs in the same vein as in the Wilczek-Zee \cite{wilczek84} and Anandan 
\cite{anandan88} approaches.  

The mixed state GP in \cite{sjoqvist00a} has been generalized and to arbitrary open system 
evolution \cite{tong04} (for some applications, see \cite{lombardo06,yi06,banerjee08,dajka11}), 
for which the probabilities of the spectral decomposition are  
generically time-dependent, i.e., $\rho (t) = \sum_{n=1}^N p_n (t) \ket{\psi_n (t)} \bra{\psi_n (t)}$ 
with $\langle \psi_n (t) \ket{\psi_m (t)} = \delta_{nm}$. The natural generalization of Eq. 
(\ref{eq:msgp_unitary}) is \cite{tong04}  
\begin{eqnarray}
\Phi [\mathcal{C}] = \arg \sum_{n=1}^N \sqrt{p_n (0) p_n (\tau)} \bra{\psi_n (0)} \psi_n (\tau) \rangle 
e^{i\int_0^{\tau} \bra{\psi_n (t)} \dot{\psi}_n (t) \rangle dt} , 
\label{eq:msgp_nonunitary}
\end{eqnarray}  
where the square root of the initial and final probablities is a consequence of 
Schmidt-form purification $\rho (t) \rightarrow \sum_{n=1}^N \sqrt{p_n (t)} \ket{\psi_n (t)} \ket{a_n}$, 
where $\{ \ket{a_n} \}$ is a fixed orthonormal ancilla basis.  

\item[(ii)] The Uhlmann GP \cite{uhlmann86,uhlmann91,hubner93}. This is a non-Abelian GP of 
mixed quantum states based on purifcation, i.e., the idea that any density operator can be written 
as the partial trace of a pure state of a larger system consisting of the considered system 
and an ancilla. This GP is defined via a parallelity condition that singles out a preferred 
purification of a density operator $\rho_{k+1}$ given a purification of another density 
operator $\rho_k$. Explicitly, a purification of a density  operator $\rho$ is a mapping 
$\rho \rightarrow \ket{\psi_{\rho}} = \sum_{n=1}^N (\sqrt{\rho} \ket{e_n}) \otimes V^{\textrm{T}} 
\ket{e_n}$, $V$ being an arbitrary unitary with $\textrm{T}$ transposition with respect to the 
fixed orthonormal basis $\ket{e_n}$. Parallelity is defined as 
\begin{eqnarray} 
\max_{V_{k+1}} \left| \langle \psi_{\rho _{k+1}} \ket{\psi_{\rho_k}} \right| = 
\Tr \sqrt{\sqrt{\rho_k} \rho _{k+1} \sqrt{\rho_k}}
\end{eqnarray}
where the right-hand side is the Uhlmann fidelity \cite{uhlmann76} that measures the 
similarity of the two states $\rho_k$ and $\rho _{k+1}$. The optimal purifications is 
given by the unitary 
\begin{eqnarray}
V_{k+1} = V_{\rho _{k+1} \rho_k} V_k , 
\end{eqnarray}
where $V_{\rho _{k+1} \rho_k}$ is the unitary part of $\sqrt{\rho _{k+1}} \sqrt{\rho_k}$. Here,   
$V_{\rho _{k+1} \rho_k}$ is a unique unitary provided the density operators are both full rank. 
The lower rank cases, such as the pure state case, must be treated separately; in these cases 
the relative phases become partial isometries. By repeating this argument along a sequence 
$\mathcal{C} : \rho_1 , \ldots , \rho_K$ we obtain the Uhlmann GP 
\begin{eqnarray}
U[\mathcal{C}] = V_{\rho_K \rho_{K-1}} \cdots V_{\rho_2 \rho_1} V_1^{\dagger} , 
\label{eq:uhlmann}
\end{eqnarray}
where the mutliplication with $V_1^{\dagger}$ from the right removes the arbitrary 
choice of initial purification, and makes the resulting unitary operator $U[\mathcal{C}]$ 
gauge invariant. The Uhlmann scheme can be applied to any continuous rank-preserving 
path by taking appropriate limit of the right-hand side of Eq. (\ref{eq:uhlmann}). 

One may notice that the above scheme is of rather mathematical nature and its physical 
meaning might seem a bit unclear. Nevertheless, it has been pointed out that it can in 
principle be realized interferometrically, provided the ancilla system can be fully controlled 
\cite{ericsson03b,aberg07}. Recent work have further shown the usefulness of the Uhlmann 
GP to analyze temperature effects in topological states of matter 
\cite{viyuela14a,huang14,viyuela14b}. 

\end{itemize}

We have seen that there are two different routes to define a GP for mixed quantum 
states. A natural question then arise: are the resulting phases in some way related? 
Various aspects of this issue have been addressed. First, it has been shown that the 
Uhlmann and interferometer GPs are indeed distinct concepts that only fully coincide 
in the special case of pure states \cite{slater02,ericsson03b}. The key difference 
lies in that while the interferometer based GP can be measured in a single-particle interference 
the Uhlmann GP is a property of a larger system in the sense that it can only be measured 
in two-particle interference \cite{ericsson03b,aberg07}. Further, the question as to 
whether the two mixed state GPs arise out of a single more fundamental notion of 
GP for mixed states has been addressed \cite{shi05,rezakhani06,andersson14}. 
 
\subsection*{\sffamily \large Physical example: Unitary evolution of a qubit} 
We illustrate the two GP concepts for mixed states in the case of unitary evolution of a 
single qubit. Any qubit density operators can be written as 
\begin{eqnarray}
\rho = \frac{1}{2} \left( \hat{1} + {\bf r} \cdot \boldsymbol{\sigma} \right) , 
\end{eqnarray}
where $\hat{1}$ is the identity operator and the Bloch vector ${\bf r}=(x,y,z)$ satisfies 
$|{\bf r}| \leq 1$, with equality if and only if $\rho$ is a pure state. Thus, the pure states 
reside on the two-dimensional Bloch sphere and the non-pure states are in the interior 
of this sphere. In particular, the origin ${\bf r} = 0$ corresponds to a equally weigthed 
mixture of any orthogonal states (random mixture). 

Let us first consider the interferometric approach. Let the eigenstates of $\rho$ be 
$\ket{\pm}$ with eigenvalues $\frac{1}{2} (1 \pm r)$, $r=|{\bf r}|$. Assume that the 
qubit undergoes cyclic unitary evolution such that the Bloch vector ${\bf r}$ traces out 
a path $\mathcal{C}$ that encloses a solid angle $\Omega$. It follows that 
$|\bra{\pm} U(\tau) \ket{\pm}| = 1$ and 
\begin{eqnarray}
|\Tr (\rho U(\tau))| e^{i\Phi [\mathcal{C}]} & = & 
\frac{1+r}{2} e^{-i\Omega /2} + \frac{1-r}{2} e^{i\Omega /2} , 
\end{eqnarray}
which yields 
\begin{eqnarray}
\Phi [\mathcal{C}] & = & -\arctan \left( r \tan \frac{\Omega}{2} \right) , 
\nonumber \\ 
|\Tr (\rho U(\tau))| & = & \sqrt{\cos^2 \frac{\Omega}{2} + r^2 \sin^2 \frac{\Omega}{2}} , 
\label{eq:msgp_qubit}
\end{eqnarray}
where the first expression gives the mixed state GP and the second is the visibility 
of the interference fringes that would be detected in an interferometer setting. Both these 
quantities are determined by the enclosed solid angle $\Omega$ on the Bloch sphere and 
the degree of mixing $r$; they reduce to the expected expressions $\Phi [\mathcal{C}] = 
-\Omega /2$ and $|\Tr (\rho U(\tau))| = 1$ in the pure state limit $r \rightarrow 1$. 
Note that the density operator is degenerate in the limit of random mixtures $r \rightarrow 0$, 
which implies that there is no unique eigenbasis of $\rho$ and the mixed state GP becomes 
undefined. Subtle interference effects close to this singularity have been examined  
\cite{bhandari02,anandan02}. 

Let us now turn to the Uhlmann GP. This GP is a non-Abelian quantity, which mean that 
it involves a time ordered product or integral. To deal with this complication, one may 
restrict to specific paths in state space. Here, we consider unitary evolution of the qubit 
around a great circle $\mathcal{G}$ inside the Bloch sphere chosen to start and end at 
the positive $z$ axis and restricted to the $xz$ plane. Such paths play a natural role to 
study temperature-driven phase transitions of the Uhlmann GP of fermion systems 
\cite{viyuela14a,huang14,viyuela14b}. We find 
\begin{eqnarray} 
U[\mathcal{G}] = - e^{i \sqrt{1-r^2} \pi \sigma_y} . 
\end{eqnarray} 
In the limit of random mixtures $r \rightarrow 0$, the Uhlmann GP is still well-defined 
but trivial, i.e., $U[\mathcal{G}] = \hat{1}$. The pure state limit $r \rightarrow 1$, on the 
other hand, is more subtle. In this limit, $\rho$ has rank 1 and $U[\mathcal{G}]$ should 
be interpreted as a U(1) phase factor times a projector. Explicitly, one finds 
\begin{eqnarray} 
U[\mathcal{G}] = - \ket{+} \bra{+} , 
\end{eqnarray}
where the minus sign coincides with the expected pure state GP of $\pi$ associated with 
a great circle on the Bloch sphere.    
  
\subsection*{\sffamily \large Measurement of mixed state GPs}  
Direct experimental tests of the mixed state GP \cite{sjoqvist00a} by using 
interferometry techniques have been carried out in NMR systems \cite{du03,ghosh06}. 
These experiments utilize two nuclear spin-$\frac{1}{2}$ with one spin playing the role 
of the interferometer arms while implementing the unitary geometric transformation on 
the second spin, conditionally on the state of the first spin. The dependence of $\Phi 
[\mathcal{C}]$ in Eq. (\ref{eq:msgp_qubit}) on the degree of mixing $r$ has been 
verified \cite{du03,ghosh06}, as well as the $r$ and $\Omega$ dependence of the visibility 
\cite{ghosh06}. A related NMR experiment \cite{cucchietti10} has examined the 
mixed state GP proposed in \cite{tong04} for non-unitary evolution. 

Experimental test of \cite{sjoqvist00a} has been performed by using polarization mixed 
states of photons in a Mach-Zehnder interferometer setup \cite{ericsson05}. These 
polarization states were prepared in two different ways, either by using birefringent decoherers 
that couple the single photon's polarization to its arrival time relative to the trigger 
or by preparing non-maximally polarization-entangled photon pairs and measuring the 
phase of one of the photons while tracing over the other. These two cases correspond to 
two types of mixed states, proper and improper ones \cite{despagnat76}, where the latter 
concept refers to that the state of the full system is in fact pure. 

In addition to interferometry, quantum mechanical phase shifts can as well be performed 
using single-beam polarimetric techniques. The basic setting to perform such experiments 
has been developed \cite{wagh95b} and later generalized to mixed states \cite{larsson03}. 
The latter has been realized in a single-beam experiment on partially polarized 
neutrons \cite{klepp05} and utilized to test the non-additive nature of phase shifts 
(geometric or dynamical) for mixed quantum states \cite{klepp08}. 

Due to the high-level of control required, the Uhlmann GP is considerably more challenging 
to implement experimentally. Nevertheless, it has been measured by using NMR technique 
for a qubit undergoing unitary evolution and comparing it with the interferometric GP for 
the same paths in the space of density operators \cite{zhu11} . The measured phases were 
explicitly shown to behave differently, providing clear experimental evidence of the 
inequivalence of the interferometer-based and Uhlmann GPs.  

\section*{\sffamily \Large ENTANGLEMENT}
Entanglement is fundamental resource in quantum information that can be used for secure 
key distribution \cite{ekert91}, teleportation \cite{bennett93}, and information processing \cite{raussendorf01}. The most basic form of entanglement is given by the four Bell states, 
which form the orthogonal basis states  
\begin{eqnarray}
\ket{\Phi_{\pm}} & = & \frac{1}{\sqrt{2}} \Big( \ket{00} \pm \ket{11} \Big) , 
\nonumber \\ 
\ket{\Psi_{\pm}} & = & \frac{1}{\sqrt{2}} \Big( \ket{01} \pm \ket{10} \Big) 
\label{eq:bell}
\end{eqnarray}
of two qubits. These are maximally entangled in the sense that each of 
them contains full information about the two qubits, but completely random information 
for each qubit separately. 

The concept of entanglement becomes even richer when considering more than two qubits. 
The additional richness lies in the fact that while two qubits essentially can be entangled in 
only one way (given by the above Bell state entanglement), there are several ways to entangle 
three or more qubits. To analyze this structure, it is convenient to use the concept of stochastic 
local operations and classical communication (SLOCC), i.e., the idea that entanglement cannot 
be changed if only local operations (unitary, measurements, etc) and classical communication 
are allowed. In this language, one may for instance show that three-qubit system in pure states can 
entangle in two different ways, in the sense that a state picked from one of these two classes 
cannot be taken to a state in the other class by performing SLOCC \cite{dur00}. The two classes, 
called $W$ and GHZ, have their basic forms 
\begin{eqnarray} 
\ket{W} & = & a\ket{001} + b\ket{010} + c\ket{100} , 
\nonumber \\ 
 \ket{\textrm{GHZ}} & = & a' \ket{000} + b'\ket{111} , 
\end{eqnarray} 
where the first is characterized by two-qubit entanglement only, while the second is genuinely 
three-qubit entangled but with no two-qubit entanglement \cite{coffman00}. 

Entanglement can be quantified by using various kinds of measures, such as entanglement 
of formation \cite{bennett96a}, geometric entanglement \cite{shimony95,wei03}, and 
relative entropy of entanglement \cite{vedral97}. Besides quantification of its amount, 
entanglement can be characterized qualitatively by using suitable polynomials of the 
expansion coefficients with respect to a product basis spanning the composite system's 
Hilbert space. For two qubits in the state $\sum_{kl} \alpha_{kl} \ket{kl}$, $\det [\alpha_{kl}] = 
\alpha_{00} \alpha_{11} - \alpha_{01} \alpha_{10}$ is a polynomial invariant under SLOCC
and therefore a property of entanglement. Furthermore, this degree-2 
polynomial determines the entanglement of formation via the concurrence measure 
\cite{wootters98} $C = 2\left| \det [\alpha_{kl}] \right|$. Similarily, the hyperdeterminant 
of the expansion coefficients $\alpha_{klm}$ of a pure three-qubit state is a SLOCC invariant 
degree-4 polynomial that determines the 3-tangle, one of the main measures of three-qubit 
entanglement \cite{coffman00}. The characterization in terms of invariant polynomials becomes increasingly complicated when the number of subsystems increases \cite{luque03,dokovic09}. 

In the following, we focus on a type of topological phases for entangled systems undergoing 
local special unitary (SU) evolution; a subclass of SLOCC operations. These topological phases 
were first discovered for two-qubit systems \cite{milman03,dezela05,liming04,milman06}, and 
later generalized to qudit ($d$-level) pairs \cite{oxman11,oxman14,khoury14} and to multi-qubit 
systems \cite{johansson12a}. They may in some cases coincide with the corresponding GPs. 
We delineate the relation between the topological phases and local SU invariant polynomials 
and describe experimental work to study these phases. The key merit of the entanglement-induced 
topological phases is that they constitute a novel perspective that may provide further insights into 
the nature of quantum entanglement. 
   
A non-exhaustive list of other studies of GP effects in entangled systems is as follows. GPs 
of bipartite qubit systems in pure states have been examined \cite{sjoqvist00b,tong03a,chen07} 
in terms of the concept of `Schmidt sphere' \cite{sjoqvist00b}, built on the analogy of the Bloch 
sphere of a single qubit, and by use of properties of braiding transformations to represent 
entangled two-qubit states \cite{chen07}. Recently, the Schmidt sphere concept has been 
examined in a quantum optical experiment \cite{loredo14}. The relation between the GPs of 
an entangled system and its subsystems has been examined \cite{tong03b,williamson07}.  
GP effects have been shown to play a role for the understanding of entanglement in multi-particle 
systems \cite{wootters04,williamson11a,williamson11b}. GPs for sequences of relative states, 
obtained by projecting on one subsystem of a bipartite composite system, have been studied 
for pure \cite{sjoqvist09} and mixed \cite{sjoqvist10} states. GPs associated with spin-orbit 
entanglement in neutron interferometry have been predicted \cite{bertlmann04} and 
experimentally observed \cite{sponar10}. A relation between the GP of a two-qubit state and 
concurrence has been delineated \cite{basu06}. Pure two-qubit states can be represented 
as single-qubit states with quaternionic expansion coefficients \cite{mosseri01}. The  
non-Abelian and entanglement-dependent GPs associated with this quaternionic representation 
have been examined \cite{levay04,johansson11}.  

\subsection*{\sffamily \large Topological phases of entangled systems: bipartite case} 
Let us first consider the case of maximally entangled states (MES) of pairs of qubits. 
Any such state can be written on the form 
\begin{eqnarray}
\ket{a,b} = \frac{1}{\sqrt{2}} \Big( a\ket{00} + b\ket{01} - b^{\ast}\ket{10} + 
a^{\ast} \ket{11} \Big) , 
\label{eq:mes}
\end{eqnarray}
where $a$ and $b$ are arbitrary complex numbers with $|a|^2 + |b|^2 = 1$. The four Bell 
states in Eq. (\ref{eq:bell}) correspond to the choices $(a,b)=(1,0),(i,0),(0,i)$, and $(0,1)$, 
respectively. These states can be transformed into each other by applying local SU(2) 
transformations $U$ and $V$. The states of the single qubit subsystems $A$ and $B$ 
are given by the reduced density operators $\rho_A$ and $\rho_B$ obtained by tracing 
over qubit $B$ and $A$, respectively. For the MES $\ket{a,b}$, the reduced density operators 
take the form $\frac{1}{2} \hat{1}_A$ and $\frac{1}{2} \hat{1}_B$, irrespective of the values of 
$a$ and $b$. Thus, the subsystem states correspond to random mixtures for all MES. 

We are interested in continuous paths $\mathcal{C}: t \in [0,\tau] \rightarrow (a(t),b(t))$ 
corresponding to the pure state local unitary evolution $\ket{a(0),b(0)} \rightarrow 
\ket{a(t),b(t)} = U(t) \otimes V(t) \ket{a(0),b(0)}$, such that the state traces 
out a loop in state space of the qubit-pair, i.e., $\ket{a(\tau),b(\tau)} = e^{if} \ket{a(0),b(0)}$. 
We can now observe two important features of this evolution: 

\begin{itemize} 

\item In order to preserve the MES form in Eq. (\ref{eq:mes}), $f$ can only take one of two 
values, namely, $0$ or $\pi$. 

\item The accumulation of local phase changes 
\begin{eqnarray} 
i\int_0^{\tau} \bra{a(t),b(t)} \frac{d}{dt} \ket{a(t),b(t)} dt = 
i \int_0^{\tau} \Tr \left( \frac{1}{2} \hat{1}_A U^{\dagger} \dot{U} \right) dt + 
i \int_0^{\tau} \Tr \left( \frac{1}{2} \hat{1}_B V^{\dagger} \dot{V} \right) dt 
\end{eqnarray} 
vanishes, since the infinitesimal generators $U^{\dagger} \dot{U}$ and 
$V^{\dagger} \dot{V}$ of $U,V\in$SU(2) are traceless. 

\end{itemize}  

\begin{figure*}[ht]
\centering
\includegraphics[width=0.8\textwidth]{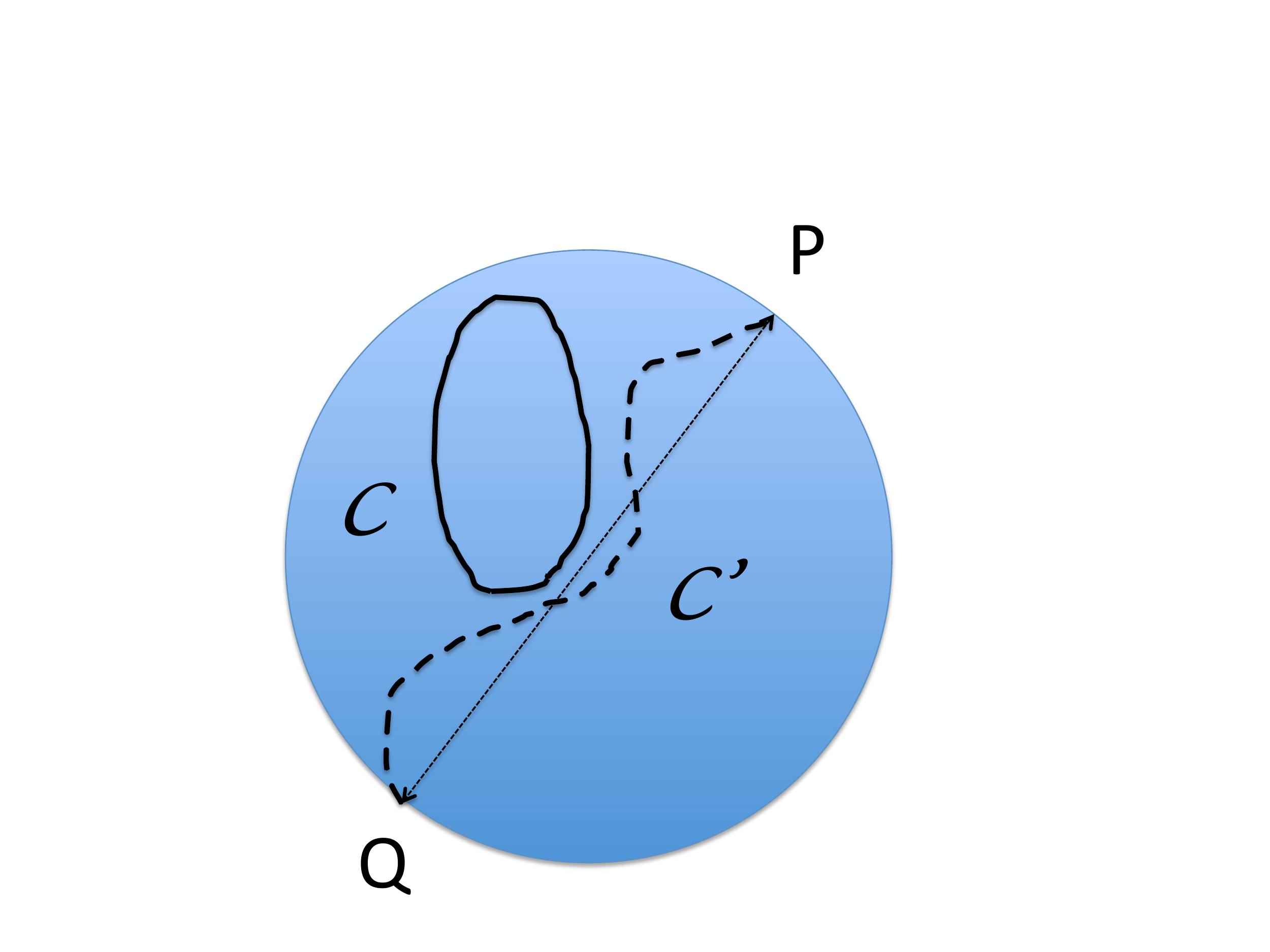}
\caption{Illustrating the two classes of topologically inequivalent loops in $S^3 /Z_2 \simeq 
\textrm{SO(3)}$. The loop $\mathcal{C}$ (solid line) can be continuously deformed 
to a point and correspond to the trivial phase factor $+1$. On the other hand, $\mathcal{C}'$ 
(dashed line) can be continuously deformed to a point only when traversed twice and 
correspond to the non-trivial phase factor $-1$. Note that $\mathcal{C}'$ is a loop since 
antipodal points on the sphere are identified, i.e., P and Q are actually the same point.} 
\label{fig5}
\end{figure*}

The first feature demonstrates the topological nature of the acquired phase $f$: it is 
impossible to transform between the two allowed values $0$ and $\pi$ by continuously 
deforming the path $\mathcal{C}$. In fact, the two phases can be associated with the 
doubly-connectedness of SO(3). This may be seen by noting that the state space of 
MES is given by  
\begin{eqnarray}
\textrm{MES} \simeq \{ (a,b) \in {C}^2 \ \textrm{such that} \ |a|^2 + |b|^2 = 1 \ \textrm{and} \ 
(a,b) \sim (-a,-b) \} .
\end{eqnarray} 
This shows that there is a one-to-one correspondence between MES and the 
three-dimensional sphere $S^3$ ($|a|^2 + |b|^2 = 1$) with antipodal points identified 
($(a,b) \sim (-a,-b)$), which is isomorphic to the three-dimensional rotation group SO(3), 
i.e., $S^3 /Z_2 \simeq \textrm{SO(3)}$. Now, SO(3) is known to be doubly-connected, 
which means that there are two types of topologically inequivalent loops, one which is 
trivial, i.e., can be continuously deformed to a point, and one which is trivial only when 
traversed twice. These two types of loops are illustrated in Figure \ref{fig5} and correspond 
to the two topological phase factors $+1$ and $-1$, respectively, first discovered by 
Milman and Mosseri \cite{milman03}. 

The second feature demonstrates that $f$ is the GP associated with the loop since the 
accumulation of local phase changes along the path vanishes. However, the local phase 
changes vanish only for maximally entangled states; for more general two-qubit states 
$\ket{\psi} = a\ket{00} + b\ket{01} + c\ket{10} + d\ket{11}$, the 
global phase factors contain also dynamical contributions, i.e.,  
\begin{eqnarray} 
i\int_0^{\tau} \langle \psi 
\ket{\dot{\psi}} dt = 
i\int_0^{\tau} \Tr \left( \rho_A U^{\dagger} \dot{U}  \right) dt + 
i \int_0^{\tau} \Tr \left( \rho_B V^{\dagger} \dot{V} \right) dt \neq 0, 
\end{eqnarray}  
since $\rho_A \neq \frac{1}{2} \hat{1}_A$ and $\rho_B \neq \frac{1}{2} \hat{1}_B$ 
if $\ket{\psi}$ is non-maximally entangled, i.e., if $2\left| ad - bc \right| < 1$. 

Note that the allowed cyclic phases $f$ are still $0$ or $\pi$ for any non-vanishing degree 
of entanglement. The only exception is when $\ket{\psi}$ is a product state, i.e., when 
$ad-bc = 0$, in case of which a continuum of cyclic phases $f$ may occur under local 
SU(2) evolution. Thus, the topological nature of the allowed global phases $f$ is intimately 
related to entanglement, in the sense that $f$ is no longer a discrete quantity exactly 
when $\ket{\psi}$ ceases to be entangled. Note that the degree-2 polynomial $ad-bc$ 
is the determinant of the expansion coefficient, which, as noted above, is known to be 
a SLOOC and thereby local SU invariant. 

The topological two-qubit phases have been generalized by Oxman and Khoury to pairs 
of $d$-level systems (qudits) \cite{oxman11} and pairs of systems with different dimension
\cite{khoury14}. For equal dimension, they found the allowed phase factors to be the 
$d$th roots of unity 
\begin{eqnarray}
e^{if} = e^{iq(2\pi/d)}, \ q=0,1,\ldots,d-1, 
\label{eq:qudit_phases}
\end{eqnarray}
and being related to topological properties of SU($d$). The above result for qubit-pairs 
is recovered as the special case where $d=2$. 

The discrete set of phases in Eq. (\ref{eq:qudit_phases}) can be understood as follows. 
Let the initial state take the general form $\ket{\psi (0)} = \sum_{k,l=1}^d \alpha_{kl} 
\ket{kl}$. Define the $d \times d$ matrix $\boldsymbol{\alpha}$ with components $\alpha_{kl}$. 
A local SU($d$) evolution $\ket{\psi (0)} \rightarrow U(t) \otimes V(t) \ket{\psi (0)}$ can be 
translated into the matrix evolution $\boldsymbol{\alpha} \mapsto \boldsymbol{u} (t) 
\boldsymbol{\alpha} {\bf v}^{\textrm{T}}(t)$, where ${\bf v} (t)$ and 
${\bf u} (t)$ are SU($d$) matrices with elements $u_{kl} (t) = \bra{k} U(t) \ket{l}$ 
and $v_{kl} (t) = \bra{k} V(t) \ket{l}$. Now, a cyclic evolution $\ket{\psi (\tau)} = 
e^{if} \ket{\psi (0)}$ yields ${\bf u} (\tau) \boldsymbol{\alpha} {\bf v}^{\textrm{T}} (\tau) = 
e^{if} \boldsymbol{\alpha}$. By taking the determinant of this expression, we obtain 
\begin{eqnarray} 
\det [{\bf u} (\tau)] \det [\boldsymbol{\alpha}] \det [{\bf v}^{\textrm{T}} (\tau)] = 
\det [\boldsymbol{\alpha}] = \left( e^{if} \right)^d \det [\boldsymbol{\alpha}] ,   
\end{eqnarray}
where we have used that $\det [{\bf u} (\tau)] = \det [{\bf v}^{\textrm{T}} (\tau)] = 1$ for 
SU($d$) matrices. If $\det [\boldsymbol{\alpha}] \neq 0$, we thus obtain that $\left( e^{if} 
\right)^d = 1$, which implies Eq. (\ref{eq:qudit_phases}).  On the other hand, if 
$\det [\boldsymbol{\alpha}] = 0$, which happens if and only if $\ket{\psi (0)}$ is a product 
state, then $f$ can take any value and is thereby no longer topological. Note that this is 
consistent with the result we found in the two-qubit case discussed above. 

\subsection*{\sffamily \large Topological phases of entangled systems: multi-qubit case} 
The topological phase structure of multi-qubit systems undergoing local SU(2) evolution  
has been examined \cite{johansson12a}. These phases show a considerably richer structure 
than those in the bipartite case, due to the richer entanglement structure of  systems 
consisting of more than two particles. 

As an illustration of this additional richness, let us consider the simplest case of three qubits 
in some detail. Here, the allowed topological phases are $0,\frac{\pi}{2},\pi,\frac{3\pi}{2}$. 
To understand the reason why just these phases, let us consider an arbitrary three-qubit 
pure state $\sum_{klm} \alpha_{klm} \ket{klm}$. The hyperdeterminant of the expansion 
coefficients $\alpha_{klm}$ takes the form  
\begin{eqnarray}
\textrm{Det} [\alpha_{klm}] & = & \alpha_{000}^2 \alpha_{111}^2 + \alpha_{001}^2 
\alpha_{110}^2 + \alpha_{010}^2 \alpha_{101}^2 + \alpha_{100}^2 \alpha_{011}^2  
\nonumber \\ 
 & & - 2\left( \alpha_{000} \alpha_{001} \alpha_{110} \alpha_{111} + \alpha_{000} 
\alpha_{010} \alpha_{101} 
\alpha_{111} + \alpha_{000} \alpha_{100} \alpha_{011} \alpha_{111} \right. 
\nonumber \\ 
 & & \left. + \alpha_{001} \alpha_{010} \alpha_{101} \alpha_{110} + 
\alpha_{001} \alpha_{100} \alpha_{011} \alpha_{110} + 
\alpha_{010} \alpha_{100} \alpha_{011} \alpha_{101} \right)  
\nonumber \\ 
 & & + 4\left( \alpha_{000} \alpha_{011} \alpha_{101} \alpha_{110} + 
\alpha_{001} \alpha_{010} \alpha_{100} \alpha_{111} \right)  , 
\end{eqnarray}  
which apparently is a degree-4 polynomial. Just as $\det [\alpha_{kl}]$ in the bipartite case, 
$\textrm{Det} [\alpha_{klm}]$ is a SLOCC invariant; in fact, it determines the 3-tangle 
$\tau_3 = 4\left|\textrm{Det} [\alpha_{klm}] \right|$. The 3-tangle measures the amount 
of genuine three-qubit entanglement in the state \cite{coffman00}. 

Now, given a cyclic local unitary evolution where all $\alpha_{klm} \rightarrow e^{if} \alpha_{klm}$,  
we obtain $\textrm{Det} [\alpha_{klm}]  \rightarrow \textrm{Det} [\alpha_{klm}] = \left( e^{if} 
\right)^4 \textrm{Det} [\alpha_{klm}]$ by using the local SU invariance of $\textrm{Det} [\alpha_{klm}]$. Thus, provided $\textrm{Det} [\alpha_{klm}] 
\neq 0$, we find $\left( e^{if} \right)^4 = 1$, which implies 
\begin{eqnarray}
f = q(2\pi/4), \ q=0,\ldots,3 . 
\end{eqnarray}
As in the bipartite case, we see that $f$ can take any value and whereby losing its topological 
nature, in cases where $\textrm{Det} [\alpha_{klm}] = 0$, which is exactly when three-qubit 
entanglement is lost (i.e., $\tau_3 = 0$). In this sense, the topological phase $f$ is a genuine 
property of three-qubit entanglement. 

By looking at the space of all pure inseparable three-qubit states, we note that the $W$-states 
is a zero measure set of this space with vanishing 3-tangle. The $W$-set constitutes a singularity 
in the space of entangled three qubits, just as the singularity line defined by the magnetic 
solenoid that gives rise to the Aharonov-Bohm topological phase \cite{aharonov59}. This further 
illustrates the toplogical nature of the allowed $f$ for non-vanishing 3-tangle. 
  
The number of allowed topological phases increases with the number of qubits. They have 
been found for up to seven qubits by using a search algorithm based on a combinatorial 
formulation of the problem \cite{johansson12a}. For eight or more qubits, only partial results 
have been established due to rapid increase in the computational resources needed to execute 
the algorithm when the number of qubits increases. 

Finally, we note that the multi-qubit phases may coincide with the corresponding GPs. As in the 
bipartite case, this happens precisely when all the marginal one-qubit states are random mixtures.  

\subsection*{\sffamily \large Experiments on topological phases of entangled systems} 
Souza {\it et al.} \cite{souza07} (see also \cite{souza14}) have used an optical setup involving 
non-separable polarization and orbital degrees of freedom to simulate quantum entanglement 
in a laser beam. The laser beam states relevant for this experiment can be described by 
multiplying orthogonal polarization unit vectors $\boldsymbol{\epsilon}_H$ and 
$\boldsymbol{\epsilon}_V$ (horizontal and vertical polarization, respectively) with a 
pair of first order Laguerre-Gaussian profiles $\psi_{\pm} ({\bf r})$ to define the laser 
beam amplitude 
\begin{eqnarray}
{\bf E} ({\bf r}) & = & \frac{1}{\sqrt{2}} \Big( \alpha \psi_+ ({\bf r}) \boldsymbol{\epsilon}_H +
\beta \psi_+ ({\bf r}) \boldsymbol{\epsilon}_V - \beta^{\ast} \psi_- ({\bf r}) \boldsymbol{\epsilon}_H + 
\alpha^{\ast} \psi_- ({\bf r}) \boldsymbol{\epsilon}_V \Big) 
\label{eq:mns}
\end{eqnarray}
representing an arbitrary maximally non-separable (MNS) mode, being a laser beam 
analog of MES for two-qubit systems. Here, $\alpha,\beta$ are complex numbers such 
that $|\alpha|^2 + |\beta|^2 = 1$. All these states are locally equivalent in the sense 
that any MNS mode can be reached by manipulating only the polarization, say, of the 
beam. This fact was utilized in \cite{souza07} to implement loops in the space of MNS 
modes by letting the laser beam pass a sequence of wave plates. In this way, the topological 
phases associated with these polarization transformations were realized and observed. 

An analogous experiment to observe the topological two-qubit phase for MES has been 
performed by using NMR technique \cite{du07}. The setup consists of two nuclear 
spin-qubits prepared in a MES and a third ancillary qubit playing the role of the two 
arms of a Mach-Zehnder type interferometer. The input state $\frac{1}{\sqrt{2}} 
(\ket{0}+\ket{1}) \otimes \ket{\textrm{MES}}$ is acted on by a conditional operation 
that takes only the MES copy connected to the $\ket{1}$ state around a loop. Thus, the 
three qubit state undergoes the transformation 
\begin{eqnarray} 
\frac{1}{\sqrt{2}} (\ket{0}+\ket{1}) \otimes \ket{\textrm{MES}} \rightarrow \frac{1}{\sqrt{2}} (\ket{0} 
\pm \ket{1}) \otimes \ket{\textrm{MES}} , 
\end{eqnarray} 
where the relative sign is precisely the topological phase factor $\pm 1$ associated 
with the loop. In this way, the topological phases were read out by measuring the 
ancilla qubit state \cite{du07}. 

The qudit and multi-qubit phases have not been verified experimentally yet. However,  
feasible experimental proposals using polarization and orbital degrees of freedom 
of photon pairs have been put forward \cite{johansson13,khoury13} and awaits to 
be performed in the future. 

\section*{\sffamily \Large CONCLUSIONS}
We have reviewed the role of geometric phase (GP) ideas in quantum information science. 
We have described how GPs can be used to perform robust quantum information processing, 
leading to the idea of holonomic or geometric quantum computation, which is a tool to realize 
quantum gates that are robust to certain errors. GP concepts for mixed quantum states 
have been described, focusing on Uhlmann's seminal work from the mid 1980s and the 
more recent development of mixed state GPs in interferometry. Finally, we have decribed 
how GPs can be used to analyze quantum entanglement, with particular focus on the 
discovery of a new type of topological phases for entangled systems.  

The importance of the reviewed body of work is two-fold. It has led to new ways to 
perform quantum information processing that may be useful to fight errors 
in quantum gate operations. This may help to reach below the error threshold, below 
which quantum error correction codes can be performed. In other words, GPs may help 
to realize large-scale quantum computation. Conversely, ideas that have been developed 
or refined within the quantum information community, such as the theory of mixed 
quantum states, open system effects, and quantum entanglement, have been applied 
to the theory of GP. In this way, further insights into the physical, mathematical, and 
conceptual nature of the GP have been obtained. 

There are a number of pertinent issues to examine in the future in this research area. 
First, explicit physical implementations of schemes that combine GQC and symmetry-aided 
error protection techniques such as decoherence free subspaces and subsystems, need to 
be developed and experimentally implemented. The key question here is whether there exist 
such schemes that at the same time avoid many-body interaction terms in the Hamiltonian 
and thereby would be much simpler to realize in the laboratory than the existing schemes. 
Secondly, the conjectured robustness of GQC compared to more conventional dynamical 
quantum gates is still an open problem that needs to be examined further. In particular, 
the robustness features of geometric gates in the presence of open system effects need  
to be addressed in realistic models that include significant non-Markovian effects.  
Thirdly, the relation between the interferometric-based and Uhlmann GPs is still not fully 
understood. A key goal here is to find a common conceptual basis for these two phases, such 
as for instance a new mixed state GP concept that covers the two. Finally, a major challenge 
concerning GP effects in entangled systems is to clarify the relation between the entanglement-induced 
topological phases and SLOCC invariant polynomials of multi-qubit systems and to 
implement experimentally these phases in the qudit and multi-qubit cases. Addressing 
the above listed open problems would lead to further insights into the nature of the GP 
and to new applications of GP ideas in quantum information science. 

\subsection*{\sffamily \large ACKNOWLEDGMENTS}
I wish to thank Jeeva Anandan, Mauritz Andersson, Vahid Azimi Mousolou, Johan Br\"annlund, 
Carlo Canali, Artur Ekert, Marie Ericsson, Bj\"orn Hessmo, Markus Johansson, Antonio Khoury, 
David Kult, Leong Chuan Kwek, Peter Larsson, Daniel Oi, Oh Choo Hiap, Arun Pati, Patrik Pawlus, 
Kuldip Singh, Jakob Spiegelberg, Anthony Sudbery, Patrik Thunstr\"om, Dianmin Tong, Vlatko Vedral, 
Mark Williamson, William Wootters, Paolo Zanardi, Guofu Xu, Jiang Zhang, and Johan {\AA}berg for 
fruitful collaboration over the years on geometric phase issues related to this review. Financial 
support from the Swedish Research Council (Vetenskapsr{\aa}det) is acknowledged. 

\end{document}